\documentclass[12pt,a4paper,final]{iopart}
\usepackage{iopams}
\usepackage{bm}

\expandafter\let\csname equation*\endcsname\relax
\expandafter\let\csname endequation*\endcsname\relax

\usepackage{amsmath}
\usepackage{amssymb}
\usepackage{amsthm}
\usepackage[utf8]{inputenc}
\usepackage{graphicx}         % Include figure files
\usepackage{bm}               % bold math
\usepackage{physics}
\usepackage{color}
\usepackage{soul}
\usepackage{cite}
\usepackage{dsfont}
\usepackage{verbatim}
\newcommand{\mycomment}[1]{}
\usepackage{mathtools}
\newtheorem{theorem}{Theorem}

\usepackage{subfig}

\let\oldproofname=\proofname
\renewcommand{\proofname}{\rm\bf{\oldproofname}} %proof en negrita

\usepackage{caption}
\DeclareCaptionFormat{custom}
{%
    \textbf{#1#2}\textit{\small #3}
}
\usepackage{xcolor}
\def\mathcolor#1#{\@mathcolor{#1}}
\def\@mathcolor#1#2#3{%
  \protect\leavevmode
  \begingroup
    \color#1{#2}#3%
  \endgroup
}

\usepackage[breaklinks=true,colorlinks=true,linkcolor=blue,urlcolor=blue,citecolor=blue]{hyperref}

\renewcommand{\eqref}[1]{Eq. (\ref{#1})}

\begin{document}

\title{A quantum expectation identity: Applications to statistical mechanics}
%\title{An useful quantum expectation identity with applications to statistical physics}

\author{Boris Maulén}
\address{Departamento de Ciencias Qu\'imicas, Facultad de Ciencias Exactas, Doctorado en Fisicoqu\'imica Molecular, Universidad Andres Bello. 8370146, Santiago, Chile.}
\address{Departamento de F\'isica, Facultad de Ciencias Exactas, Universidad Andres Bello. 8370136, Santiago, Chile.}
\ead{b.maulenjara@uandresbello.edu}

\author{Sergio Davis}
\address{Research Center in the intersection of Plasma Physics, Matter and Complexity (P$^2$mc), Comisi\'on Chilena de Energía Nuclear, Casilla 188-D, Santiago, Chile.}
\address{Departamento de F\'isica, Facultad de Ciencias Exactas, Universidad Andres Bello. Sazi\'e 2212, piso 7, 8370136, Santiago, Chile.}

\author{Daniel Pons}
\address{Departamento de Matemáticas, Facultad de Ciencias Exactas,  Universidad Andres Bello, Santiago, Chile.}

\begin{abstract}
In this article we derive a useful expectation identity using the language of quantum statistical mechanics, where density matrices represent the state of knowledge about the system. This identity allows to establish relations between different quantum observables depending on a continuous parameter $\gamma \in \mathbb{R}$. Such a parameter can be contained in the observables itself (e.g. perturbative parameter) or may appear as a Lagrange multiplier (inverse temperature, chemical potential, etc.) in the density matrix, excluding parameters that modify the underlying Hilbert space. In this way, using both canonical and grand canonical density matrices along with certain quantum observables (Hamiltonian, number operator, the density matrix itself, etc.) we found new identities in the field, showing not only its derivation but also their meaning. Additionally, we found that some theorems of traditional quantum statistics and quantum chemistry, such as the thermodynamical fluctuation-dissipation theorem, the Ehrenfest, and the Hellmann-Feynman theorems, among others, are particular instances of our aforementioned quantum expectation identity. At last, using a \textit{generalized} density matrix arising from the Maximum-Entropy principle, we derive generalized quantum expectation identities: these generalized identities allow us to group all the previous cases in a unitary scheme.
\end{abstract}

\section{Introduction}

%\nocite{Davis2012}

The concept of expectation value plays a central role both in statistical sciences and physics. Within the subjective point of view of probabilities \cite{jaynes1}, it corresponds to the prediction that a theory makes on a variable or a certain observable of a system under study. Also, the quality of a prediction depends directly on how much information we have about the variable under consideration. For example, if we have not only the expectation value of a variable but also certain statistical moments (e.g., variance), our knowledge will be more complete ( but not necessarily more precise). In this sense, an expectation value accompanied by a small variance will lead to a good prediction. In contrast, an expectation value accompanied by a large variance will lead to a poor prediction. 
Examples of expectation values in physics are the ensemble average in classical statistical mechanics, the pure-state mean value of basic quantum mechanics, and the mixed-state mean value of quantum statistics \cite{greiner}. In order to give a number to an expectation value, it is necessary to count with a statistical distribution of the accessible states of the respective physical system. For example, to calculate the ensemble average of a phase-space function we only need to integrate that function weighted by a phase-space distribution over all coordinates and all the classical momenta. However, in several cases, the resulting integral can not be evaluated straightforwardly. 

On the other hand, following the Bayesian formulation of statistics \cite{bayesian_stat_mech}, it is possible to derive an algebraic identity that relates the derivative of an expectation value taken in a certain state of knowledge with another expectation taken in the same state of knowledge. Specifically, for an arbitrary function $\omega(\mathbf{X};\gamma)$ of a random variable $\mathbf{X}$, which depends parametrically on a continuous parameter $\gamma$ and follows a nowhere vanishing statistical distribution $P(\mathbf{X}|\gamma,I)$, we have
\begin{equation}
\dfrac{\partial}{\partial \gamma} \left\langle \omega  \right\rangle_{\gamma, I}=\int_U d\mathbf{x} \left[ \dfrac{\partial \omega(\mathbf{x};\gamma)}{\partial \gamma} + \omega(\mathbf{x};\gamma) \dfrac{\partial \ln P(\mathbf{X}=\mathbf{x}|\gamma,I)}{\partial \gamma}  \right]P(\mathbf{X}=\mathbf{x}|\gamma,I), \label{classical_FDT} 
\end{equation} 
where $\gamma$ is also a parameter contained in the statistical distribution, $U$ is the support of the density $P(\mathbf{X}|\gamma,I)$ such as $\mathbf{x}\in U$, and $I$ points out a certain state of knowledge. It is worth noting that in Eq. ($\ref{classical_FDT}$), the expectation values are taken not only in the state of knowledge given by $I$, but also by the continuous parameter $\gamma$. The identity shown in Eq. (\ref{classical_FDT}) is a type of classical expectation identity in which the variation (derivative) is taken outside the expectation \cite{divergence}. With the aid of Eq. (\ref{classical_FDT}), it is possible, for example, to evaluate indirectly the mean value and the variance associated with the discrete variable $k$ in the Poisson distribution without the need to evaluate any integral \cite{divergence}. Applied to classical statistical mechanics, Eq. (\ref{classical_FDT}) allows us to recover certain identities such as the relationship between the mean energy with the logarithmic derivative of the canonical partition function and the relationship between the specific heat of a thermodynamical system with its energy fluctuation: this relation corresponds to the thermodynamical version of the fluctuation-dissipation theorem (FDT)\cite{kardar}.

In the case of quantum mechanics, we have two types of expectation values depending on whether we are dealing with pure quantum states or mixed quantum states. In the first case, we use the wave function $\psi_n$ associated with a certain quantum number $n$ as a probability amplitude, to calculate the expectation value of a Hermitian operator, while in the case of mixed states, it is necessary to know the density matrix $\hat{\rho}$ of the system. Thus, for mixed-state quantum systems, the density matrix becomes a key quantity since it, being the equivalent of the phase-space distribution in quantum mechanics, acts as a probability distribution of the accessible quantum states.

Depending on the type of the system, we have different methodologies to obtain the density matrix $\hat{\rho}$ associated with the quantum state of such a system. For example, in atomic and molecular systems, we can reconstruct the density matrix using some fundamental parameters of the respective model, such as the total charge and the number of electrons, and also imposing well-known constraints to the statistical weights (probabilities) \cite{bochicchio_maulen}. On the other hand, within the inferential statistical approach of E. T. Jaynes \cite{jaynes1, jaynes_book}, we can build the least biased quantum distribution only maximizing the von-Neumann entropy  
\begin{equation}
S=-k\Tr \left\lbrace   \hat{\rho}\log \hat{\rho} \right\rbrace  , \label{von_neumann_entropy}
\end{equation}
under specific constraints related to whatever we know about the system (in Eq. (\ref{von_neumann_entropy}) $k$ is the Boltzmann constant). This procedure is, indeed, the application of the Maximum Entropy Principle (MaxEnt) to quantum statistics \cite{jaynes2, maxent_rdm}. In practice, the constraints that are used in the MaxEnt procedure are given by different Lagrange multipliers, such as the inverse temperature $\beta$, or the product $\alpha=-\beta\mu$, where $\mu$ is the chemical potential, etc. In this sense each Lagrange multiplier is associated with a respective expectation value of a key observable of the system: in the case of $\beta$ the mean energy, in the case of $\alpha$ the mean particle number. Additionally, the knowledge of the expectation values of quantum observables allows us to reconstruct the density matrix of a microscopic system. Such a process is known as Quantum State Tomography \cite{tomography, efficient_reconstruction_tomography, tomography_reduced_density}, which, nowadays, comprises an important ground to quantum computing \cite{computing}.

Once we have the density matrix of a system, we only need a basis in order to expand the trace operator in the expression
\begin{equation}
\left\langle \hat{A} \right\rangle_{\hat{\rho}}=\Tr \left\lbrace \hat{A} \hat{\rho} \right\rbrace, \label{expectation_value}
\end{equation}
where $\hat{A}=\hat{A}^{\dag}$ is a Hermitian linear operator acting on the Hilbert space $\mathcal{H}$ of the accessible states of the system. It is important to stress that we have used the density matrix $\hat{\rho}$ as a subscript in the expectation value (\ref{expectation_value}) because, in quantum statistics and following the Bayesian nomenclature, it represents the state of knowledge in which we are calculating the expectation of $\hat{A}$. If the density matrix $\hat{\rho}$ used in Eq. (\ref{expectation_value}) comes from the MaxEnt principle, it may depend on a collection of Lagrange multipliers, so that, the expectation value (\ref{expectation_value}) depends on the same parameters too. We will call that collection of Lagrange multipliers \textbf{statistical-like} parameters since these determine the form of the statistical model given by the density matrix. Moreover, there are cases in which the observables of the system may depend on extra parameters, different from those already mentioned, such as an oscillation frequency, a perturbative parameter, time, etc. We will call such a collection of parameters \textbf{quantum-like} parameters because these are only contained in the quantum observables (e.g. the Hamiltonian) and never appear outside of them.   

The main goal of this article is to adapt Eq. (\ref{classical_FDT}) to the quantum world, taking special care in the way how the concepts are translated from classical to quantum statistics, and to show that several traditional theorems of quantum statistics and quantum mechanics, such as the Hellmann-Feynman and the Ehrenfest theorems, the thermodynamical FDT, and the generalization of the thermodynamic integration identity (useful in determining the Helmholtz free energy difference in numerical simulations) among others, are different instances of our \textbf{quantum expectation identity} (QEI, see Eq. (\ref{FD_theorem}) below). Finally, by means of a \textit{generalized} density matrix coming from the MaxEnt principle we obtain generalized quantum expectation identities: we will conclude that all of the previous identities, such as the canonical and the grand canonical identities, are particular instances of such generalized relationships. 

The above suggests the following reflection: \textit{Are the aforementioned theorems (Ehrenfest theorem, thermodynamical FDT, etc.) physical relationships or, on the contrary, are only straightforward applications of the Bayesian reasoning to quantum statistical mechanics?} \cite{quantum_probabilities,quantum_mechanics_quantum_information,causally_neutral_bayesian,qbism_locality}.  Without
giving a definitive answer, we, at least, will seed the question.

This article is organized as follows: Section \ref{derivation_section} is devoted to the statement of the theorem and the corresponding proof of our QEI. In Section \ref{hellmann-feynman} we review the Hellmann-Feynman theorem as a particular case of our quantum identity, alluding to its importance in quantum chemistry and molecular physics. Section \ref{ehrenfest_section} provides a view of the Ehrenfest theorem from our QEI. Section \ref{canonical_grand-canonical_section} is dedicated to reviewing applications to the canonical and grand canonical models of quantum statistical mechanics. In Section \ref{purity_section} we show a novel application of the QEI in the context of quantum information theory when obtaining differential equations for the quantum purity from our QEI. At last Section \ref{generalized_section} is devoted to the generalized quantum identities coming from a MaxEnt density matrix. 

A remark about the notation: in this work we use the symbol $\ln$ for the natural logarithm of a scalar function, and the symbol $\log$  for the natural logarithm of an operator. The circumflex accent or \textit{hat} over a symbol $A$, namely $\hat{A}$, denotes an  operator on the underlying Hilbert space, as is customary in the quantum chemistry and chemical physics literature \cite{levine}.

%By last, in section \textbf{6} we derive FD identities coming from an arbitrary function $F(\hat{\rho})$ of the density matrix, exploring the particular cases in which such a function is the density matrix itself or is the \textit{entropy-estimator}, i.e., $F(\hat{\rho})=\hat{\rho}$ and $F(\hat{\rho})=-\log \hat{\rho}$.
%
%
%
\section{Theorem and proof} \label{derivation_section}
Consider an observable $\hat{A}(\gamma)$ and a mixed quantum state represented by a density matrix $\hat{\rho}(\gamma)$, acting both on a Hilbert space $\mathcal{H}$ with a fixed domain $\Omega$. Both observables are dependent on a continuous parameter $\gamma \in \mathbb{R}$ which can be of the following types:
\begin{itemize}
\item[$\mathbf{1}$)] \textit{Quantum-like parameters}: Are parameters that can be contained in the Hamiltonian $\hat{H}(\gamma)$ or other observables of the system, such as a perturbative parameter, time or some coordinate.

\item[$\mathbf{2}$)] \textit{Statistical-like parameters}: Are parameters that are not contained in the observables of the system, and appear as Lagrange multipliers associated with some constraint. Examples of this type of parameters are the inverse temperature $\beta$, chemical potential $\mu$, etc.
\end{itemize}
In this article, we exclude parameters that change the underlying Hilbert space. For example, those that change the size or shape of the domain $\Omega$ of the system (such as the length of the well in the  model of the potential-well) are not considered in this work.

We are interested in calculating the derivative of the expectation value of $\hat{A}(\gamma)$ in the state $\hat{\rho}(\gamma)$ with respect to the parameter $\gamma$, i.e.
\begin{equation}
\dfrac{\partial}{\partial \gamma} \left\langle \hat{A}(\gamma) \right\rangle_{\hat{\rho}(\gamma)}=\dfrac{\partial}{\partial \gamma} \Tr \left\lbrace \hat{A}(\gamma) \hat{\rho}(\gamma) \right\rbrace. \label{derivation-1}
\end{equation}

For the sake of clarity, it is convenient to have in mind two important concepts that will be used in the statement of Theorem 1, namely, \textit{compatible observables} and \textit{non-singularity}:

\begin{itemize}
\item[$i)$] For two given observables $\hat{A}$ and $\hat{B}$, they are said to be \textit{compatible} if the respective operators acting on the underlying Hilbert space commute, i.e. $\left[\hat{A},\hat{B} \right]=0 $.

\item[$ii)$] An operator $\hat{C}$ is said to be $\textit{non-singular}$ if it has an inverse denoted by $\hat{C}^{-1}$, such that 
\begin{equation}
\nonumber \hat{C}\hat{C}^{-1} = \hat{C}^{-1}\hat{C}=\mathds{1},
\end{equation} 
where $\mathds{1}$ is the identity operator.

\end{itemize}

With the above considerations we enunciate the following theorem:
\begin{theorem} \textbf{(Quantum Expectation Identity)} \label{theorem_1}

Let $\hat{A}(\gamma)$ be a quantum observable and $\hat{\rho}(\gamma)$ be the density matrix of the system, both dependent on the parameter $\gamma \in \mathbb{R}$, and acting on the underlying Hilbert space $\mathcal{H}$. Suppose that:

\begin{itemize}

\item[1)] $\hat{A}(\gamma)$ and $\hat{\rho}(\gamma)$ are compatible observables. 

\item[2)] $\hat{\rho}(\gamma)$ is a non-singular operator. 

\end{itemize}
We conclude
\begin{equation}
\boxed{\dfrac{\partial}{\partial \gamma} \left\langle \hat{A}(\gamma) \right\rangle_{\hat{\rho}(\gamma)}=\left\langle \dfrac{\partial \hat{A}(\gamma) }{\partial\gamma} \right\rangle_{\hat{\rho}(\gamma)}+\left\langle \hat{A}(\gamma)\dfrac{\partial  \log \hat{\rho}(\gamma) }{\partial\gamma}  \right\rangle_{\hat{\rho}(\gamma)}.}\label{FD_theorem}
\end{equation}

\end{theorem}
\begin{proof}

Due to the property of the invariance of the expectation with respect to the basis, we are free to choose, in Eq. (\ref{derivation-1}), a continuous or a discrete basis kets. Here we consider, as is usual in quantum mechanics, that the Hilbert space is \textit{separable}, i.e. there is at least one dense countable sequence in $\mathcal{H}$ \cite{velarde}. Hence, in order to expand the trace operator in (\ref{derivation-1}), we use the eigenstates of $\hat{\rho}(\gamma)$, and, due to the above-mentioned property of separability of $\mathcal{H}$, we can assume that they form a countable basis $\left\lbrace \ket{n;\gamma} \right\rbrace_{n \in \mathbb{N}}$ labeled both by a quantum number $n\in \mathbb{N}$ and by the parameter $\gamma$. 

We can write
\begin{equation}
\begin{split}
\dfrac{\partial}{\partial \gamma} \Tr \left\lbrace \hat{A}(\gamma) \hat{\rho}(\gamma) \right\rbrace & =\dfrac{\partial }{\partial \gamma} \sum_{n \in \mathbb{N}} \bra{n;\gamma} \hat{A}(\gamma) \hat{\rho}(\gamma) \ket{n;\gamma}\\
& = \sum_{n \in \mathbb{N}} \left[ \dfrac{\partial}{\partial\gamma} \bra{n;\gamma}  \right] \hat{A}(\gamma) \hat{\rho}(\gamma) \ket{n;\gamma}\\
& +\sum_{n \in \mathbb{N}} \bra{n;\gamma} \hat{A}(\gamma) \hat{\rho}(\gamma) \left[ \dfrac{\partial}{\partial \gamma} \ket{n;\gamma}  \right]\\
& +\sum_{n \in \mathbb{N}} \bra{n;\gamma} \dfrac{\partial}{\partial \gamma} \left[   \hat{A}(\gamma) \hat{\rho}(\gamma) \right]   \ket{n;\gamma}. \label{FD1}
\end{split}
\end{equation}
Now, according to the Hypothesis ($1$), we have  $\left[ \hat{A},\hat{\rho} \right]=0$, hence $\hat{A}$ and $\hat{\rho}$ have the common eigenbasis $\left\lbrace \ket{n;\gamma} \right\rbrace_{n \in \mathbb{N}}$. The above is also true even when the density matrix $\hat{\rho}(\gamma)$ has degeneracies. For example, suppose that $\hat{\rho}(\gamma)$ has, for a given eigenvalue, say $\rho_n$, an $s$-fold degeneracy. Then, there are $s$ different states spanning the subspace $\mathcal{H}_n \subset \mathcal{H}$ that give the same probability $\rho_n$ after a measurement of the observable $\hat{\rho}(\gamma)$. In such a case, to differentiate between all quantum states of $\hat{\rho}$ that give the same eigenvalue $\rho_n$, we would need an additional quantum number to label the common eigenbasis of $\hat{\rho}(\gamma)$ and $\hat{A}(\gamma)$, namely $\left\lbrace \ket{n,i;\gamma}\right\rbrace$, with $n\in\mathbb{N}$ and $i=1,2,...,s$.\footnote[1]{An analogous situation appears in the context of quantum angular momentum. The squared angular momentum operator $\mathbf{\hat{J}}^2$ and the $z$-component of the angular momentum $\hat{J}_z$, are compatible observables. However, there are $2j+1$ different eigenstates of $\hat{J}_z$, labeled by $m=-j,-j+1,...,0,...,j-1,j$, linked to the same value of the squared angular momentum, namely $\hbar^2 j(j+1)$. Therefore, a common eigenbasis of $\mathbf{\hat{J}}^2$ and $\hat{J}_z$ must be labeled by both $j$ and $m$: $\left\lbrace \ket{j,m} \right\rbrace_{j\in \mathbb{N}}$, $m=0,\pm 1, \pm 2,...,\pm j$  \cite{sakurai}.} Nevertheless, for the sake of simplicity, we will carry out this proof considering that there are no degenerate probabilities, but keeping in mind what the case would be if this were to occur. Based on the above considerations, the spectral decompositions of $\hat{A}$ and $\hat{\rho}$ are given by
\begin{equation}
\begin{split}
& \hat{A}(\gamma)=\sum_{n \in \mathbb{N}}a_n(\gamma)\hat{P}_n(\gamma),\\
& \hat{\rho}(\gamma)=\sum_{n \in \mathbb{N}}\rho_n(\gamma)\hat{P}_n(\gamma),
\end{split}
\end{equation}
where $a_n(\gamma)$ and $\rho_n(\gamma)$ are the eigenvalues of $\hat{A}$ and $\hat{\rho}$, respectively. Moreover, $\rho_n(\gamma)$ represents the associated probabilities to each pure state $\hat{P}_n(\gamma)=\ket{n;\gamma}\bra{n;\gamma}$ to which the system has access, and whose form depends on the particular statistical model (e.g. canonical, grand canonical, etc.). Knowing the eigenvalues of $\hat{A}(\gamma)$ and $\hat{\rho}(\gamma)$, we can use them in the first two sums of the right-hand side of $\left( \ref{FD1} \right)$, thus obtaining
\begin{equation}
\begin{split}
& \left[ \dfrac{\partial}{\partial \gamma}  \bra{n;\gamma} \right]  \hat{A}(\gamma) \hat{\rho}(\gamma)\ket{n;\gamma}+\bra{n;\gamma} \hat{A}(\gamma) \hat{\rho}(\gamma) \left[ \dfrac{\partial}{\partial \gamma} \ket{n;\gamma}  \right]  \\
& = \rho_n(\gamma) \left\lbrace \left[ \dfrac{\partial}{\partial \gamma}  \bra{n;\gamma} \right] \hat{A}(\gamma)\ket{n;\gamma} + \bra{n;\gamma} \hat{A}(\gamma) \left[ \dfrac{\partial}{\partial \gamma} \ket{n;\gamma}  \right] \right\rbrace \\
& = \rho_n(\gamma) a_n(\gamma) \left\lbrace \left[ \dfrac{\partial}{\partial \gamma}  \bra{n;\gamma} \right] \ket{n;\gamma} + \bra{n;\gamma} \left[ \dfrac{\partial}{\partial \gamma} \ket{n;\gamma}  \right] \right\rbrace,
\end{split}
\end{equation}
and assuming normalized states $\ket{n;\gamma}$, i.e. $\braket{n;\gamma}=1$, we recall that
\begin{equation}
\dfrac{\partial  }{\partial \gamma} \left[ \braket{n;\gamma}\right] = \left\lbrace \left[ \dfrac{\partial}{\partial \gamma}  \bra{n;\gamma} \right] \ket{n;\gamma} + \bra{n;\gamma} \left[ \dfrac{\partial}{\partial \gamma} \ket{n;\gamma}  \right] \right\rbrace=0. \label{orthonormalization_states}
\end{equation}
Thus, Eq. $\left( \ref{FD1} \right)$ is reduced to
\begin{equation}
\dfrac{\partial}{\partial\gamma} \left\langle \hat{A}(\gamma) \right\rangle_{\hat{\rho}(\gamma)}= \sum_{n \in \mathbb{N}} \bra{n;\gamma} \dfrac{\partial \hat{A}(\gamma) }{\partial\gamma} \hat{\rho}(\gamma) \ket {n;\gamma} + \sum_{n \in \mathbb{N}} \bra{n;\gamma} \hat{A}(\gamma) \dfrac{\partial \hat{\rho}(\gamma) }{\partial\gamma}  \ket {n;\gamma}, \label{FD2}
\end{equation}
where the first sum in Eq. ($\ref{FD2}$) is easily related to the expectation of $\partial \hat{A}/\partial \gamma$ in the state $\hat{\rho}(\gamma)$,
\begin{equation}
\sum_{n \in \mathbb{N}} \bra{n;\gamma} \dfrac{\partial \hat{A}(\gamma) }{\partial\gamma} \hat{\rho}(\gamma) \ket {n;\gamma}=\left\langle \dfrac{\partial \hat{A}(\gamma) }{\partial\gamma} \right\rangle_{\hat{\rho}(\gamma)}. 
\end{equation}
In order to rewrite as an expectation value the second sum in Eq. (\ref{FD2}), we need to assume that $\hat{\rho}$ is a non-singular operator (Hypothesis (2)). This can always be done by discarding all possible zero eigenvalues of $\hat{\rho}$ and considering only those that are non-zero (i.e. non-zero probabilities). Then, multiplying by $\mathds{1}=\hat{\rho}\hat{\rho}^{-1}$ between $\hat{A}$ and $\partial \hat{\rho}/\partial \gamma$ in the second sum of Eq. (\ref{FD2}) and using Hypothesis (1) ($[\hat{A},\hat{\rho}]=0$), we have
\begin{equation}
\begin{split}
\sum_{n \in \mathbb{N}} \bra{n;\gamma} \hat{A} \dfrac{\partial  \hat{\rho} }{\partial\gamma} \ket {n;\gamma} & =\sum_{n \in \mathbb{N}} \bra{n;\gamma} \hat{A}\hat{\rho}\hat{\rho}^{-1}\dfrac{\partial  \hat{\rho} }{\partial\gamma} \ket {n;\gamma}\\ 
& = \sum_{n \in \mathbb{N}} \bra{n;\gamma} \hat{\rho} \hat{A}\hat{\rho}^{-1}\dfrac{\partial  \hat{\rho} }{\partial\gamma} \ket {n;\gamma}\\
& = \left\langle \hat{A} \hat{\rho}^{-1}\dfrac{\partial  \hat{\rho} }{\partial\gamma}   \right\rangle_{\hat{\rho}}  \label{FD3}
\end{split}
\end{equation}
On the other hand, in the Appendix we sketch a proof of the equality
\begin{equation}
\hat{\rho}^{-1}\dfrac{\partial  \hat{\rho}}{\partial\gamma} =\dfrac{\partial \log \hat{\rho}}{\partial \gamma},\label{log-rho}
\end{equation}
which follows from the Taylor series of $\hat{\rho}^{-1}$ and $\log \hat{\rho}$. With the above we can recast Eq. (\ref{FD3}) by use of (\ref{log-rho}), namely
\begin{equation}
\left\langle \hat{A}(\gamma) \hat{\rho}^{-1}(\gamma)\dfrac{\partial  \hat{\rho}(\gamma) }{\partial\gamma}   \right\rangle_{\hat{\rho}(\gamma)}= \left\langle \hat{A}(\gamma) \dfrac{\partial \log \hat{\rho} (\gamma)}{\partial \gamma}  \right\rangle_{\hat{\rho}(\gamma)}.
\end{equation}
Therefore, we conclude that the derivative with respect to a continuous parameter $\gamma$ of the expectation value of an observable $\hat{A}(\gamma)$ is given by
\begin{equation}
\dfrac{\partial}{\partial \gamma} \left\langle \hat{A}(\gamma) \right\rangle_{\hat{\rho}(\gamma)}=\left\langle \dfrac{\partial \hat{A}(\gamma) }{\partial\gamma} \right\rangle_{\hat{\rho}(\gamma)}+\left\langle \hat{A}(\gamma)\dfrac{\partial  \log \hat{\rho}(\gamma) }{\partial\gamma}  \right\rangle_{\hat{\rho}(\gamma)}.
\end{equation}

\end{proof}
Theorem (\ref{theorem_1}) allows to relate the derivative of an expectation value with the sum of two other expectations taken in the same state of knowledge given by the density matrix $\hat{\rho}(\gamma)$. The conclusion of Theorem (\ref{theorem_1}), Eq. ($\ref{FD_theorem}$), is the quantum version of Eq. (\ref{classical_FDT}) for observables compatible with the density matrix, and corresponds to our aforementioned \textbf{quantum expectation identity} (QEI). By means of Eq. (\ref{FD_theorem}) we will derive new identities in the field (see Sections \ref{grand_canonical_section} and \ref{purity_section} below) and also the following standard results of quantum mechanics and thermodynamics (among others):
\begin{itemize}
\item[a)] The Hellmann-Feynman theorem.
\item[b)] The Ehrenfest theorem.
\item[c)] The relation between the partition function and the mean energy of a quantum system.
\item[d)] The thermodynamic integration formula.
\item[e)] The thermodynamical version of the fluctuation-dissipation theorem.
\end{itemize}

At last, it is important to remark that a simple way to ensure that  $\left[ \hat{A},\hat{\rho} \right]=0$, is when both $\hat{A}$ and $\hat{\rho}$ are functions of the Hamiltonian $\hat{H}$, i.e. $\hat{A}=A( \hat{H}) $ and $\hat{\rho}=\rho( \hat{H})$.

\section{Hellmann-Feynman from QEI} \label{hellmann-feynman}
In the context of classical statistics, the derivative with respect to a continuous parameter $\gamma$ on the expectation value of a function $\omega(\mathbf{X};\gamma)$ in the state of knowledge given only by $I$, is
\begin{equation}
\dfrac{\partial}{\partial \gamma} \left\langle \omega(\mathbf{X};\gamma) \right\rangle_{I} = \left\langle \dfrac{\partial \omega(\mathbf{X};\gamma)}{\partial \gamma}  \right\rangle_{I}. \label{derivative_expectation_value}
\end{equation}
We observe that Eq. (\ref{derivative_expectation_value}) is a particular case of Eq. (\ref{classical_FDT}): the last reduces to the former in the case when the statistical distribution $P(\mathbf{X}|I)$ does not depend on the parameter $\gamma$ used in the derivative. 

In quantum mechanics, we have an analogous situation. Consider the QEI Eq. (\ref{FD_theorem}) for an operator $\hat{A}(\gamma)$ and for the instance where the density matrix $\hat{\rho}$ represents a mixed state but does not contain the parameter $\gamma$ used in the derivative. In such a case, the QEI reduces to
\begin{equation}
\dfrac{\partial}{\partial \gamma} \left\langle \hat{A}(\gamma) \right\rangle_{\hat{\rho}}=\left\langle \dfrac{\partial \hat{A}(\gamma) }{\partial\gamma} \right\rangle_{\hat{\rho}}. 
\end{equation} 
Observe that the above identity is the quantum equivalent of the Eq. (\ref{derivative_expectation_value}). Then, in the special case in which the operator is the Hamiltonian of the system, i.e. $\hat{A}(\gamma)=\hat{H}(\gamma)$, where $\gamma$ may be some parameter contained in it (mass, force constant, etc.), we have
\begin{equation}
\dfrac{\partial }{\partial \gamma} \left\langle \hat{H} (\gamma) \right\rangle_{\hat{\rho}}= \left\langle \dfrac{\partial \hat{H}(\gamma)}{\partial \gamma} \right\rangle_{\hat{\rho}}. \label{Hellmann-Feynman-mixed}
\end{equation}
Eq. (\ref{Hellmann-Feynman-mixed}) can be considered as an instance of the Hellmann-Feynman theorem (see Eq. (\ref{Hellmann-Feynman-pure}) below) for the case where the quantum state is not described by a single wave function, but by a mixed state $\hat{\rho}$.

Consider now the case where $\hat{A}(\gamma)=\hat{H}(\gamma)$ as before, and where $\hat{\rho}$ represents a pure state projector, namely $\hat{\rho}=\ket{n;\gamma}\bra{n;\gamma}$ (which may depend on $\gamma$). We have
\begin{equation}
\begin{split}
\dfrac{\partial}{\partial \gamma} \left\langle \hat{H}(\gamma) \right\rangle_{n} & =\dfrac{\partial}{\partial \gamma}\Tr{\hat{H}(\gamma)\ket{n;\gamma}\bra{n;\gamma}} \\
& = \dfrac{\partial}{\partial \gamma} \sum_{m\in \mathbb{N}} \bra{m;\gamma}\hat{H}(\gamma)\ket{n;\gamma} \delta_{n,m}\\
& = \dfrac{\partial}{\partial \gamma} \left[ \bra{n;\gamma} \hat{H}(\gamma)\ket{n;\gamma} \right] \\
& = \bra{n;\gamma} \dfrac{\partial \hat{H}(\gamma)}{\partial \gamma} \ket{n;\gamma},
\end{split}
\end{equation}
where we have assumed that $ \left\lbrace \ket{n;\gamma}\right\rbrace _{n\in \mathbb{N}}$ is an orthonormal eigenbasis of $\hat{H}(\gamma)$ (cf. Eq. (\ref{orthonormalization_states})). Then, considering that the expectation value of $\hat{H}(\gamma)$ taken in the pure state $\ket{n;\gamma}$ gives the energy eigenvalue $E_n(\gamma)$, 
\begin{equation}
\left\langle \hat{H}(\gamma) \right\rangle_{n} = \bra{n;\gamma} \hat{H}(\gamma)\ket{n;\gamma}=E_n(\gamma),
\end{equation}
we finally obtain
\begin{equation}
\dfrac{\partial E_n(\gamma)}{\partial \gamma}= \left\langle  \dfrac{\partial \hat{H}(\gamma)}{\partial \gamma}\right\rangle_n. \label{Hellmann-Feynman-pure}
\end{equation}
Eq. (\ref{Hellmann-Feynman-pure}), which is called the \textit{Hellmann-Feynman} theorem, is widely used in the context of quantum chemistry and molecular physics to explore the chemical bonding in molecular systems \cite{levine, jensen}.

%Not only the Hellmann-Feynman theorem is recovered when the density matrix $\hat{\rho}$ does not depend on $\gamma$, but also when it represents a pure state projector, namely $\hat{\rho}=\ket{n;\gamma}\bra{n;\gamma}$ (which may depend on $\gamma$). In this case, we have

An interesting use of the Hellmann-Feynman theorem, in the context of quantum chemistry, is in the case in which the parameter $\gamma$ corresponds to a certain nuclear coordinate $R_{i,\alpha}$ ($ith$ coordinate of the nucleus $\alpha$) of a $N$-electron molecular system. Considering the molecular Schr\"odinger equation within the Born-Oppenheimer approximation \cite{levine}
\begin{equation}
\left[ \hat{H}_{el}+\hat{V}_{NN}   \right] \ket{\psi_{el}}=U(\mathbf{R})\ket{\psi_{el}},
\end{equation} 
where $\hat{H}_{el}=\hat{T}_{el}+\hat{V}_{el}$ is the \textit{electronic} Hamiltonian (which includes the electronic kinetic energy operator $\hat{T}_{el}$ and the electronic potential operator $\hat{V}_{el}$), $\hat{V}_{NN}$ is the potential due to nuclear repulsion and $\ket{\psi_{el}}$ are the electronic states of the molecule. The energy $U(\mathbf{R})$ corresponds to the electronic energy plus the internuclear repulsion and is called the \textit{Born-Oppenheimer energy} (BO) of the molecule \cite{borkman_parr}. We notice that the BO energy depends parametrically on the coordinates of the nuclei in the molecule throughout the nuclear coordinate vector $\mathbf{R}=\left(R_{x,1},R_{y,1},R_{z,1},...,R_{x,M},R_{y,M},R_{z,M}\right)$, where $M$ is the total number of nuclei in the molecule. In such a case, the Hellmann-Feynman theorem can be written as (for $\gamma=R_{i,\alpha}$)
\begin{equation}
\dfrac{\partial U(\mathbf{R})}{\partial R_{i,\alpha}} = \int d\mathbf{r}_1...d\mathbf{r}_{N}\hspace{0.1cm} \psi_{el}^{*}(\mathbf{r}_1,...,\mathbf{r}_{N})\left[ \dfrac{\partial V_{el}}{\partial R_{i,\alpha} } +\dfrac{\partial V_{NN}}{\partial R_{i,\alpha} }  \right]   \psi_{el}(\mathbf{r}_1,...,\mathbf{r}_{N}), 
\end{equation}
where $\mathbf{r}_i$ denotes the coordinates of the $ith$ electron. The derivative of the BO energy with respect to $R_{i,\alpha}$
can be seen as the $ith$ component of a force on the $\alpha th$ nucleus due to the other nuclei and the electron cloud. See \cite{Davidson_book} for more details.

\section{Ehrenfest theorem from QEI} \label{ehrenfest_section}
An important result of quantum mechanics is the Eherenfest theorem, which states that \cite{sakurai}
\begin{equation}
\dfrac{d }{dt} \left\langle \hat{A} \right\rangle_{\ket{\psi}} = \left\langle  \dfrac{\partial \hat{A}}{\partial t} \right\rangle_{\ket{\psi}} +\dfrac{1}{i\hbar} \left\langle \left[ \hat{A},\hat{H} \right]  \right\rangle_{\ket{\psi}}, \label{ehrenfest_theorem} 
\end{equation}
and accounts for the time evolution of the expectation value of an observable $\hat{A}$ in a closed quantum system with a Hamiltonian $\hat{H}$. Originally, the expectations in Eq. (\ref{ehrenfest_theorem}) are taken in a pure state $\ket{\psi}\bra{\psi}$, but this can easily be generalized to the case of mixed states.

In order to derive the Ehrenfest theorem (Eq. (\ref{ehrenfest_theorem})) from the QEI, we use in Eq. (\ref{FD_theorem}) an arbitrary time-dependent density matrix $\hat{\rho}(t)$ with an arbitrary time-dependent observable  $\hat{A}(t)$,
\begin{equation}
\dfrac{d }{dt} \left\langle \hat{A}(t) \right\rangle_{\hat{\rho}(t)}=\left\langle \dfrac{\partial \hat{A}(t)}{\partial t} \right\rangle_{\hat{\rho}(t)}  + \left\langle \hat{A}(t) \dfrac{\partial \log \hat{\rho}(t)}{ \partial t}  \right\rangle_{\hat{\rho}(t)}, \label{ehrenfest1}
\end{equation}
where we have chosen the time $t$ as the parameter $\gamma$. Consider the second term on the right-hand side of Eq. (\ref{ehrenfest1}) and use $\left[\hat{A},\hat{\rho} \right]=0$ (Hypothesis (1)) and $\hat{\rho}^{-1}\hat{\rho}=\mathds{1}$ in order to rewrite it as
\begin{equation}
%\begin{split}
\left\langle \hat{A}(t) \dfrac{\partial \log \hat{\rho}(t)}{ \partial t}  \right\rangle_{\hat{\rho}(t)}  =\left\langle \hat{A}(t)  \hat{\rho}^{-1}(t)  \dfrac{\partial \hat{\rho}(t)}{ \partial t}   \right\rangle_{\hat{\rho}(t)} =\Tr \left\lbrace \hat{A}(t) \dfrac{\partial \hat{\rho}(t)}{\partial t} \right\rbrace.
%\end{split}
\end{equation} 
Assuming also that $\hat{\rho}(t)$ evolves in an unitary way, i.e., according to the Liouville-von Neumann equation of motion
\begin{equation}
\dfrac{\partial \hat{\rho}(t)}{\partial t}=\dfrac{1}{i\hbar}\left[\hat{H}(t), \hat{\rho}(t) \right], \label{von_neumann}
\end{equation}
we obtain
\begin{equation}
\left\langle \hat{A}(t) \dfrac{\partial \log \hat{\rho}(t)}{ \partial t}  \right\rangle_{\hat{\rho}(t)}=\dfrac{1}{i\hbar}\Tr \left\lbrace \hat{A}(t) \left[ \hat{H}(t),\hat{\rho}(t) \right]  \right\rbrace. \label{ehrenfest2}
\end{equation} 
After some algebra, it is possible to recast the trace operator in (\ref{ehrenfest2}) as an expectation value in the state $\hat{\rho}(t)$,
\begin{equation}
\Tr \left\lbrace \hat{A} \left[ \hat{H},\hat{\rho} \right] \right\rbrace = \left\langle \left[ \hat{A},\hat{H}\right]  \right\rangle_{\hat{\rho}(t)}, 
\end{equation}
where we have used the cyclic property of the trace operator. Finally, the last expression allows us to recover the Ehrenfest theorem from the QEI
\begin{equation}
\dfrac{d }{dt} \left\langle \hat{A}(t) \right\rangle_{\hat{\rho}(t)}=\left\langle \dfrac{\partial \hat{A}(t)}{\partial t} \right\rangle_{\hat{\rho}(t)}+\dfrac{1}{i\hbar} \left\langle \left[ \hat{A}(t),\hat{H}(t)\right]  \right\rangle_{\hat{\rho}(t)}.
\end{equation}

\section{Canonical and grand canonical identities} \label{canonical_grand-canonical_section}
Before giving the general results obtained from the QEI for a generalized density matrix, we illustrate the use of ($\ref{FD_theorem}$) in the canonical and grand canonical statistical models. According to Section \ref{derivation_section}, we are free to choose $\hat{A}$ as any observable compatible with $\hat{\rho}$ such as the identity operator $\mathds{1}$, the Hamiltonian $\hat{H}$ and the density matrix $\hat{\rho}$ itself, or even an arbitrary function $F(\hat{\rho})$ of $\hat{\rho}$. For the sake of simplicity, in this section we consider only the the cases $\hat{A}=\mathds{1}$ and $\hat{A}=\hat{H}$. We will consider the case $\hat{A}=\hat{\rho}$ in Section \ref{purity_section} separately.

%, and we leave for the section $\mathbf{6}$ the general case $\hat{A}=F(\hat{\rho})$ where the chosen observable is an arbitrary function of the density matrix.

\subsection{Canonical model} \label{canonical_model_section}
We consider a canonical density matrix of the form
\begin{equation}
\hat{\rho}_c(\beta)=\dfrac{\exp{-\beta \hat{H}}}{\mathcal{Z}(\beta)}, \label{canonical_density_matrix}
\end{equation}
where $\beta$ is the inverse temperature and
\begin{equation}
\mathcal{Z}(\beta)=\Tr\left( e^{-\beta\hat{H}}\right)
\end{equation}
is the canonical partition function, which depends on $\beta$ and, in certain cases, on a quantum-like parameter $\lambda$. If we use the canonical model Eq. (\ref{canonical_density_matrix}) in the QEI, we obtain
\begin{equation}
\dfrac{\partial}{\partial \gamma} \left< \hat{A} \right>_{\hat{\rho}_c} = \left< \dfrac{\partial \hat{A}}{\partial \gamma} \right>_{\hat{\rho}_c}-\left<\hat{A} \dfrac{\partial}{\partial \gamma} \left( \beta \hat{H}  \right)   \right>_{\hat{\rho}_c}- \dfrac{\partial \ln\mathcal{Z}(\beta)}{\partial \gamma}\left< \hat{A} \right>_{\hat{\rho}_c}, \label{canonical_qfdt}
\end{equation}
where $\gamma$ may be a quantum-like ($\lambda$) or a statistical-like parameter ($\beta$). In the following, we will use in the above relation Eq. (\ref{canonical_qfdt})  (which we have called the \textit{canonical QEI}) $\hat{A}=\mathds{1}$ and $\hat{A}=\hat{H}$ for both types of parameters, obtaining four canonical identities (which are summarized in Table \ref{table_1}):

\begin{itemize}
\item[$i)$] \textbf{$\hat{A}=\mathds{1}$ and $\gamma$ is a quantum-like parameter ($\lambda$)}:

In this case, we have that
\begin{equation}
\nonumber \dfrac{\partial}{\partial \lambda} \left\langle  \mathds{1}  \right\rangle=0 \hspace{0.5cm} \text{and}\hspace{0.5cm}\left\langle \dfrac{\partial  \mathds{1}}{\partial \lambda} \right\rangle=0,
\end{equation}
so the second term on the right-hand side of Eq. (\ref{FD_theorem}) becomes 
\begin{equation}
\begin{split}
0=\left\langle \dfrac{\partial \log \hat{\rho}_c}{\partial \lambda}  \right\rangle_{\hat{\rho}_c} & = \left\langle \dfrac{\partial}{\partial \lambda} \left[-\beta \hat{H}(\lambda) -\ln\mathcal{Z}(\beta,\lambda)\mathds{1} \right] \right\rangle_{\hat{\rho}_c}\\
& = -\beta \left\langle \dfrac{\partial \hat{H}(\lambda)}{\partial \lambda} \right\rangle_{\hat{\rho}_c}-\dfrac{\partial \ln\mathcal{Z}(\beta,\lambda)}{\partial \lambda},
\end{split}
\end{equation}
and we obtain
\begin{equation}
\left\langle \dfrac{\partial \hat{H}(\lambda)}{\partial \lambda}\right\rangle _{\hat{\rho}_c}=-\dfrac{1}{\beta}\dfrac{\partial \ln \mathcal{Z}(\beta,\lambda)}{\partial \lambda}\bigg|_{\beta}.  \label{canonical1}
\end{equation}
Identifying in the above identity the Helmholtz Free energy,
\begin{equation}
\mathcal{F}=-\dfrac{1}{\beta}\ln \mathcal{Z}(\beta,\lambda), \label{free_energy}
\end{equation}
we can rewrite this (Eq. (\ref{canonical1})) in the form
\begin{equation}
\left\langle \dfrac{\partial \hat{H}(\lambda)}{\partial \lambda}\right\rangle _{\hat{\rho}_c}=\dfrac{\partial \mathcal{F}(\beta,\lambda)}{\partial \lambda}. 
\end{equation}
Finally, we can obtain the free energy difference $\Delta\mathcal{F}$ integrating with respect to $\lambda \in \left[ \lambda_{min},\lambda_{max} \right]$ on both sides:

\begin{equation}
\Delta \mathcal{F}=\int_{\lambda_{min}}^{\lambda_{max}}d\lambda \left\langle \dfrac{\partial \hat{H}(\lambda)}{\partial \lambda}\right\rangle _{\hat{\rho}_c}.
\end{equation}

This procedure is used to obtain free energy differences via molecular dynamics, such as the ionic hydration free energy difference, and is called the \textit{thermodynamic integration technique} \cite{free_energy,free_energy2}.

\item[$ii)$] \textbf{$\hat{A}=\mathds{1}$ and $\gamma$ is a statistical-like parameter ($\beta$)}:

As in the previous case, we have
\begin{equation}
\nonumber \dfrac{\partial }{\partial \beta} \left\langle \mathds{1} \right\rangle = 0 \hspace{0.5cm}\text{and}\hspace{0.5cm}\left\langle  \dfrac{\partial \mathds{1}}{\partial \beta} \right\rangle=0, 
\end{equation} 
and the QEI gives
\begin{equation}
\begin{split}
0=\left\langle \dfrac{\partial \log \hat{\rho}_c}{\partial \beta}  \right\rangle_{\hat{\rho}_c} &=\left\langle \dfrac{\partial}{\partial \beta} \left[ -\beta \hat{H}-\ln \mathcal{Z}(\beta) \mathds{1} \right] \right\rangle_{\hat{\rho}_c}\\
& =- \left\langle \hat{H} \right\rangle_{\hat{\rho}_c}-\dfrac{\partial \ln \mathcal{Z}(\beta)}{\partial \beta},
\end{split}
\end{equation}
where we have used the fact that $\hat{H}$ does not depend on the inverse temperature $\beta$. In this way, we get
\begin{equation}
\left\langle \hat{H}  \right\rangle_{\hat{\rho}_c}=-\dfrac{\partial \ln \mathcal{Z}(\beta)}{\partial \beta}\bigg|_{\lambda}. \label{canonical2}
\end{equation}
It is important to observe that the identity shown in Eq. (\ref{canonical2}) is the well-known relationship between the partition function and the average energy which arises in the context of canonical ensemble in traditional statistical mechanics \cite{greiner}.

\item[$iii)$] \textbf{$\hat{A}=\hat{H}$ and $\gamma$ is a quantum-like parameter ($\lambda$)}:

In this case we have
\begin{equation}
\begin{split}
\dfrac{\partial }{\partial \lambda} \left\langle \hat{H} (\lambda) \right\rangle_{\hat{\rho}_c}& =\left\langle \dfrac{\partial \hat{H}(\lambda) }{\partial \lambda} \right\rangle_{\hat{\rho}_c}+\left\langle \hat{H}(\lambda) \dfrac{\partial }{\partial \lambda} \left[ -\beta \hat{H}(\lambda) -\ln\mathcal{Z}(\beta,\lambda)\mathds{1} \right]  \right\rangle_{\hat{\rho}_c}   \\
& = \left\langle \dfrac{\partial \hat{H}(\lambda) }{\partial \lambda} \right\rangle_{\hat{\rho}_c} -\beta \left\langle \hat{H}(\lambda)\dfrac{\partial \hat{H}(\lambda)}{\partial \lambda}\right\rangle_{\hat{\rho}_c}  -\dfrac{\partial \ln \mathcal{Z}(\beta,\lambda)}{\partial \lambda} \left\langle \hat{H}(\lambda) \right\rangle_{\hat{\rho}_c}, 
\end{split} 
\end{equation}

and using Eq. (\ref{canonical1}), we finally get
\begin{equation}
\dfrac{\partial }{\partial \lambda} \left\langle \hat{H} (\lambda) \right\rangle_{\hat{\rho}_c} =\left\langle \dfrac{\partial \hat{H}(\lambda) }{\partial \lambda} \right\rangle_{\hat{\rho}_c}-\beta \hspace{0.05cm} \mathrm{Cov}\left(\hat{H}(\lambda),\dfrac{\partial \hat{H}(\lambda)}{\partial \lambda} \right), \label{canonical_H_lambda}
\end{equation}
where $\mathrm{Cov}\left( \hat{H},\dfrac{\partial \hat{H}}{\partial \lambda} \right)$ stands for
\begin{equation}
\begin{split}
\mathrm{Cov}\left( \hat{H},\dfrac{\partial \hat{H}}{\partial \lambda} \right) & \equiv \left<  \Delta\hat{H} \hspace{0.05cm}\Delta \hspace{-0.1cm}\left( \dfrac{\partial \hat{H}}{\partial \lambda} \right) \right>_{\hat{\rho}_c} \\
& = \left\langle \hat{H} \dfrac{\partial \hat{H}}{\partial \lambda} \right\rangle_{\hat{\rho}_c}-\left\langle \hat{H} \right\rangle _{\hat{\rho}_c} \left\langle \dfrac{\partial \hat{H}}{\partial \lambda} \right\rangle_{\hat{\rho}_c}, \label{cov}
\end{split}  
\end{equation} 
i.e., corresponds to the covariance between the Hamiltonian and its derivative in the state of knowledge given by $\hat{\rho}$, and measures the correlation between both operators in a statistical sense. In Eq. (\ref{cov}) $\Delta \hat{H}$ means the deviation of $\hat{H}$ from its expectation (mean) value,
\begin{equation}
\nonumber \Delta\hat{H}= \hat{H}-\left< \hat{H} \right> \mathds{1},
\end{equation}
and the same for $\partial \hat{H} / \partial \lambda$.

It is worth mentioning that two compatible observables (i.e. whose commutator is null), might not be uncorrelated. This can be seen if we recast the covariance between any two observables $\hat{A}$ and $\hat{B}$ as \cite{sakurai}
\begin{equation}
\mathrm{Cov}\left( \hat{A},\hat{B} \right)=\dfrac{1}{2} \left< \left[\hat{A},\hat{B}\right] \right>_{\hat{\rho}}+\dfrac{1}{2}\left< \left\lbrace  \Delta\hat{A}, \Delta \hat{B}  \right\rbrace \right>_{\hat{\rho}}, \label{cov2}
\end{equation} 
where 
\begin{equation}
\nonumber \left\lbrace \Delta\hat{A}  , \Delta\hat{B}  \right\rbrace = \Delta\hat{A} \Delta\hat{B}+\Delta\hat{B}\Delta\hat{A}
\end{equation}
stands for the anticommutator between the deviations of $\hat{A}$ and $\hat{B}$. In this way, we can easily see that if $\left[ \hat{A},\hat{B} \right]=0$, the anticommutator term survives and so does the covariance.

\item[$iv)$] \textbf{$\hat{A}=\hat{H}$ and \textcolor{red}{$\gamma$} is a statistical-like parameter ($\beta$)}: 

In this case, the QEI gives
\begin{equation}
\begin{split}
\dfrac{\partial}{\partial \beta} \left\langle \hat{H}  \right\rangle_{\hat{\rho}_c} & = \left\langle  \hat{H} \dfrac{\partial}{\partial \beta} \left[ -\beta \hat{H}-\ln\mathcal{Z}(\beta)\mathds{1} \right]  \right\rangle_{\hat{\rho}_c}\\
& = -\left\langle \hat{H}^2 \right\rangle_{\hat{\rho}_c}-\dfrac{\partial \ln \mathcal{Z}(\beta)}{\partial \beta} \left\langle  \hat{H} \ \right\rangle_{\hat{\rho}_c} \label{canonical3}
\end{split}
\end{equation}
where we have used that $\partial \hat{H}/\partial \beta =0$.

Using Eq. (\ref{canonical2}) in the second term of Eq. (\ref{canonical3}), we finally obtain
\begin{equation}
\dfrac{\partial}{\partial \beta} \left\langle \hat{H} \right\rangle_{\hat{\rho}_c}  = -\mathrm{Var}\left( \hat{H} \right), \label{canonical4}
\end{equation}
where $\mathrm{Var} \left( \hat{H} \right)$ is the variance in the energy of the system, namely
\begin{equation}
\mathrm{Var} \left( \hat{H} \right) = \left\langle \hat{H}^2\right\rangle_{\hat{\rho}_c}  - \left\langle \hat{H} \right\rangle^2_{\hat{\rho}_c}.  
\end{equation}
Then, by means of the chain rule and using $\beta=1/kT$, we finally obtain
\begin{equation}
C=\dfrac{\partial}{\partial T} \left\langle \hat{H} \right\rangle_{\hat{\rho}_c}=\dfrac{1}{kT^2} \mathrm{Var}\left( \hat{H} \right), \label{FD_termo} 
\end{equation}
where $C$ is the heat capacity of the system, $T$ is the absolute temperature and $k$ is the Boltzmann constant. Traditionally, Eq. (\ref{FD_termo}) is understood as one instance of the fluctuation-dissipation theorem (FDT) \cite{kardar} (thermodynamical FDT); it relates an external change in energy, driven by a change in temperature (hence, dissipation) with the internal fluctuations of the energy in equilibrium. 

%It supports the name given to Eq. (\ref{FD_theorem}), namely the fluctuation-dissipation theorem.

\end{itemize}

\begin{center}
\begin{table} 
\begin{footnotesize}
\centering
\begin{tabular}{| c || c | c  |} 
\hline 
\textbf{Observable} & \color{blue}$\gamma=\lambda$ (quantum-like parameter) & \color{red}$\gamma=\beta$ \color{black}(statistical-like parameter) \\ \hline                        
$\hat{A}=\mathds{1}$ & $\left\langle \dfrac{\partial \hat{H}(\lambda)}{\partial \lambda}\right\rangle=-\dfrac{1}{\beta}\dfrac{\partial \ln \mathcal{Z}(\beta,\lambda)}{\partial \lambda}\bigg|_{\beta}$  &  $\left\langle \hat{H}  \right\rangle=-\dfrac{\partial \ln \mathcal{Z}(\beta)}{\partial \beta}\bigg|_{\lambda}$ \\
$\hat{A}=\hat{H}$  & $\dfrac{\partial }{\partial \lambda} \left\langle \hat{H} (\lambda) \right\rangle =\left\langle \dfrac{\partial \hat{H}(\lambda) }{\partial \lambda} \right\rangle-\beta \hspace{0.05cm} \mathrm{Cov}\left(\hat{H}(\lambda),\dfrac{\partial \hat{H}(\lambda)}{\partial \lambda} \right)$   & $\dfrac{\partial}{\partial \beta} \left\langle \hat{H} \right\rangle  = -\mathrm{Var}\left( \hat{H} \right)$  \\ 
\hline  
\end{tabular}
\caption{Canonical quantum expectation identities} \label{table_1}
\end{footnotesize}
\end{table}
\end{center}

\subsection{Grand canonical model} \label{grand_canonical_section}

The grand canonical density matrix corresponds to the following statistical model:
\begin{equation}
\hat{\rho}_g(\beta,\mathit{z})=\dfrac{\exp\left\lbrace -\beta\left( \hat{H}-\mu \hat{N}  \right) \right\rbrace }{\Xi(\beta,\mathit{z})}, \label{grand_canonical_density_matrix}
\end{equation}
where $\mu$ is the chemical potential, $\hat{N}$ is the number operator and $\Xi(\beta,\mathit{z})$ is the grand partition function which depends on the inverse temperature $\beta$ and the \textit{fugacity} $z=e^{\beta\mu}$. As in the canonical model, the grand partition function may also depend on a quantum-like parameter $\lambda$.

Using the grand canonical model Eq. (\ref{grand_canonical_density_matrix}) in the QEI (\ref{FD_theorem}), we obtain the \textit{grand canonical QEI}:
\begin{small}

\begin{equation}
\dfrac{\partial }{\partial \gamma} \left< \hat{A} \right>_{\hat{\rho}_g} = \left< \dfrac{\partial \hat{A}}{\partial \gamma}  \right>_{\hat{\rho}_g}-\left< \hat{A}\dfrac{\partial}{\partial \gamma} \left( \beta \hat{H} \right)   \right>_{\hat{\rho}_g}+\left< \hat{A} \dfrac{\partial}{\partial \gamma} \left( \beta \mu \hat{N} \right)  \right>_{\hat{\rho}_g}-\dfrac{\partial \ln\Xi(\beta,\mathit{z})}{\partial \gamma}\left< \hat{A} \right>_{\hat{\rho}_g},\label{gc_qei}
\end{equation}
\end{small}
where, in the present case, $\gamma$ may be a quantum-like ($\lambda$) or a statistical-like parameter ($\beta$ or $\mu$). It is important to note that, in the case where the derivative is taken with respect to the inverse temperature, the fugacity is fixed. In such a case, the third term in Eq. (\ref{gc_qei}), because $\hat{N}$ is a constant operator, does not contribute.

As above, we can replace $\hat{A}$ with certain observables compatibles with the grand canonical density matrix $\hat{\rho}_g$. In this case we have the following three options:
\begin{itemize}
\item[\textbf{1})] $\hat{A}=\mathds{1}$
\item[\textbf{2})] $\hat{A}=\hat{H}$
\item[\textbf{3})] $\hat{A}=\hat{N}$
\end{itemize}  
We obtain for the grand canonical statistical model nine different quantum expectation identities, which are summarized in Table \ref{table_2}. Some of these identities are well-known relations coming from quantum statistics while others are new ones. At this point, as mentioned above, the number operator $\hat{N}$ is, in general, a constant operator, so that a quantum-like parameter $\lambda$ can only appear within the Hamiltonian. Since the calculations concerning the grand canonical model follow in a very similar way to the calculations of the canonical model, here we will omit the details.

Considering the first row in Table \ref{table_2} for $\hat{A}=\mathds{1}$. We have that these three cases (using $\gamma=\lambda$, $\gamma=\beta$ and $\gamma=\mu$) lead to known relationships of quantum statistical mechanics. In particular for the pair ($\mathds{1},\lambda$) we obtain 
\begin{equation}
\left<   \dfrac{\partial \hat{H}}{\partial \lambda} \right>_{\hat{\rho}_g(\lambda)}= -\dfrac{1}{\beta} \dfrac{\partial \ln\Xi(\beta,\mathit{z},\lambda)}{\partial \lambda}\bigg|_{\beta,\mathit{z}}, \label{grand_canonical0}
\end{equation}
where the derivative has been taken fixing the fugacity and the inverse temperature. As in the canonical model, we can identify another thermodynamic potential: the grand potential $\Phi(\beta,\mathit{z},\lambda)$. It is defined by 
\begin{equation}
\Phi(\beta,\mathit{z},\lambda)=-\dfrac{1}{\beta}\ln\Xi(\beta,\mathit{z},\lambda).
\end{equation}
If we integrate Eq. (\ref{grand_canonical0}) with respect to $\lambda\in \left[\lambda_{min},\lambda_{max} \right]$, using the definition of $\Phi$, we can obtain the grand potential difference:
\begin{equation}
\Delta \Phi(\beta,\mathit{z},\lambda)=\int_{\lambda_{min}}^{\lambda_{max}} d\lambda \left<   \dfrac{\partial \hat{H}}{\partial \lambda} \right>_{\hat{\rho}_g}. \label{grand_potential_difference}
\end{equation}  
Eq. (\ref{grand_potential_difference}) is another instance of the thermodynamic integration technique applied to open quantum systems. Then, for the pairs ($\mathds{1},\beta$) and $(\mathds{1},\mu)$, we have
\begin{equation}
\left< \hat{H} \right>_{\hat{\rho}_g}=-\dfrac{\partial \ln \Xi(\beta,\mathit{z})}{\partial \beta}\bigg|_{\lambda,\mathit{z}} \label{grand_canonical1}
\end{equation}
and
\begin{equation}
\left< \hat{N} \right>_{\hat{\rho}_g}=\dfrac{1}{\beta} \dfrac{\partial \ln \Xi(\beta,\mathit{z})}{\partial \mu}\bigg|_{\lambda,\beta}, \label{grand_canonical2}
\end{equation}
respectively. The identities in Eqs. (\ref{grand_canonical1}) and (\ref{grand_canonical2}), which relate, in the first case, the grand partition function with the mean energy for an open quantum system, and in the second case, the grand partition function with the mean particle number, are well-known identities in quantum statistics \cite{greiner}. It is important to highlight that the derivative in Eq. (\ref{grand_canonical1}) has been taken fixing both $\lambda$ and $\mathit{z}$, while in Eq. (\ref{grand_canonical2}) both $\lambda$ and $\beta$ are fixed.

Consider now the second row in Table \ref{table_2} for $\hat{A}=\hat{H}$.  For the pairs $(\hat{H},\lambda)$ and $(\hat{H},\beta)$ we obtain the same identities (Eqs. (\ref{canonical_H_lambda}) and (\ref{canonical4})) as for the canonical model, but using  $\hat{\rho}_g$ in the calculations of the mean values. Because we have the same identities for both canonical and grand canonical models, one may be tempted to interpret this as a consequence of the principle of equivalence between different statistical ensembles. However, such a principle is meaningless in our case, because we are interested in small quantum systems in which it is not possible to employ the \textit{thermodynamic limit} ($N\rightarrow\infty$): such equivalence between these canonical and grand canonical identities is an algebraic consequence of the similitude between both statistical models; the grand canonical model can be considered as the canonical model with an extra condition. 

Next, for the pair $(\hat{H},\mu)$ we have the identity
\begin{equation}
\dfrac{\partial}{\partial \mu} \left<  \hat{H} \right>_{\hat{\rho}_g}=\beta \hspace{0.05cm} \mathrm{Cov}\left( \hat{H},\hat{N}\right), \label{grand_canonical_H_mu}
\end{equation}
which accounts for the dependence of the variation of the mean energy with respect to the chemical potential expressed as the covariance between the Hamiltonian and the number operator: the stronger the correlation between $\hat{H}$ and $\hat{N}$, the greater dependence of the energy with respect to the chemical potential.

Consider finally the third row of Table \ref{table_2} for $\hat{A}=\hat{N}$. As above, we obtain three relations, namely for $\gamma=\lambda$, $\gamma=\beta$ and $\gamma=\mu$. The first two cases ($\gamma=\lambda$ and $\gamma=\beta$), along with Eq. (\ref{grand_canonical_H_mu}), to the best of our knowledge, are new identities. Thus, for the pair ($\hat{N},\lambda$) we obtain
\begin{equation}
\dfrac{\partial}{\partial \lambda} \left<  \hat{N} \right>_{\hat{\rho}_g}=-\beta\hspace{0.05cm}\mathrm{Cov}\left( \hat{N},\dfrac{\partial \hat{H}}{\partial \lambda} \right);
\end{equation}
identity that has been derived using Eq. (\ref{grand_canonical0}). Then, for the pair ($\hat{N},\beta$), we obtain
\begin{equation}
\dfrac{\partial}{\partial \beta}\left< \hat{N} \right>_{\hat{\rho}_g}=-\mathrm{Cov}\left( \hat{N},\hat{H}\right),\label{grand_canonical_N_beta}
\end{equation}
where we have used Eq. (\ref{grand_canonical1}). As in Eq. (\ref{grand_canonical_H_mu}), the correlation between the Hamiltonian and the number operator allows to predict the change between a property of the system (the mean particle number) driven by the change in a external parameter (the inverse temperature). At last, for the pair ($\hat{N},\mu$), and by means of Eq. (\ref{grand_canonical2}), the following identity is obtained:

\begin{equation}
\dfrac{\partial}{\partial \mu} \left<  \hat{N} \right>_{\hat{\rho}_g}=\beta \hspace{0.05cm}\mathrm{Var}\left( \hat{N}\right). \label{grand_canonical_N_mu}
\end{equation}
Eq. (\ref{grand_canonical_N_mu}), which can be derived also in a traditional way by applying derivatives with respect to the chemical potential on the grand partition function, allows to predict the change in the mean particle number with an external parameter (chemical potential) only measuring the internal fluctuations in the particle number. For this reason, we can interpret Eq. (\ref{grand_canonical_N_mu}) as another instance of the thermodynamical FDT, but for the conjugate-variable couple $(\hat{N},\mu)$. In this article we call to the relations such as Eqs. (\ref{FD_termo}), (\ref{grand_canonical_H_mu}), (\ref{grand_canonical_N_beta}) and (\ref{grand_canonical_N_mu}) fluctuation-dissipation-like identities.

%...................................................................................................................................

\begin{table}
\begin{scriptsize}
\centering
\begin{tabular}{| c || c | c | c |}
\hline 
\textbf{Obs.} & \color{blue} $\gamma=\lambda$  & \color{red}$\mathbf{\gamma=\beta}$  & \color{orange} $\mathbf{\gamma=\mu}$ \\ \hline                        
$\hat{A}=\mathds{1}$ & $\left<   \dfrac{\partial \hat{H}}{\partial \lambda} \right>= -\dfrac{1}{\beta} \dfrac{\partial \ln\Xi(\beta,\mathit{z},\lambda)}{\partial \lambda}\bigg|_{\beta,\mathit{z}}$ & $\left< \hat{H} \right>=-\dfrac{\partial \ln \Xi(\beta,\mathit{z})}{\partial \beta}\bigg|_{\lambda,\mathit{z}}$ & $\left< \hat{N} \right>=\dfrac{1}{\beta} \dfrac{\partial \ln \Xi(\beta,\mathit{z})}{\partial \mu}\bigg|_{\lambda,\beta}$ \\

$\hat{A}=\hat{H}$    & $\dfrac{\partial }{\partial \lambda} \left\langle \hat{H}  \right\rangle =\left\langle \dfrac{\partial \hat{H} }{\partial \lambda} \right\rangle-\beta \hspace{0.05cm} \mathrm{Cov}\left(\hat{H},\dfrac{\partial \hat{H}}{\partial \lambda} \right)$ & $\dfrac{\partial }{\partial \beta} \left\langle  \hat{H}\right\rangle=-\mathrm{Var} \left( \hat{H} \right)$ & $\dfrac{\partial}{\partial \mu} \left<  \hat{H} \right>=\beta \hspace{0.05cm} \mathrm{Cov}\left( \hat{H},\hat{N}\right)$ \\

$\hat{A}=\hat{N}$    & $\dfrac{\partial}{\partial \lambda} \left<  \hat{N} \right>=-\beta\hspace{0.05cm}\mathrm{Cov}\left( \hat{N},\dfrac{\partial \hat{H}}{\partial \lambda} \right)$  & $\dfrac{\partial}{\partial \beta}\left< \hat{N} \right>=-\mathrm{Cov}\left( \hat{N},\hat{H}\right)$  &  $\dfrac{\partial}{\partial \mu} \left<  \hat{N} \right>=\beta \hspace{0.05cm}\mathrm{Var}\left( \hat{N}\right)$ \\
\hline  
\end{tabular}
\caption{Grand canonical quantum expectation identities.} \label{table_2}
\end{scriptsize}
\end{table}

\section{Quantum purity and the QEI} \label{purity_section}
As already mentioned in the introduction of this article, there are two types of quantum states, namely, the pure ones and the mixed ones. A way to quantify how pure is a quantum state is introducing the quantity 
\begin{equation}
\varphi \equiv \Tr\left\lbrace \hat{\rho}^2\right\rbrace 
\end{equation}
called \textit{purity} associated with the quantum state $\hat{\rho}$ \cite{wilde}. Consider first a pure state represented by a single projector $\hat{\rho}_{pure}=\ket{\psi}\bra{\psi}$. It is worth noting that the density matrix of a pure state is idempotent, i.e.
\begin{equation}
\hat{\rho}^2_{pure}=\ket{\psi}\braket{\psi}\bra{\psi}=\ket{\psi}\bra{\psi}=\hat{\rho}_{pure}.
\end{equation}
Using this property, the purity of $\hat{\rho}_{pure}$ is
\begin{equation}
\varphi\left( \hat{\rho}_{pure} \right) = \Tr \left\lbrace \hat{\rho}^2_{pure}  \right\rbrace= \Tr\left\lbrace \hat{\rho}_{pure}  \right\rbrace=1,
\end{equation}
where the last equality follows from the normalization of the density matrix. On the other hand, the purity associated with a mixed state is strictly less than $1$, and tends to zero as the dimensionality of the Hilbert space increases \cite{purity_pennini}.

A novel application of the QEI in the context of quantum information theory arises when using the density matrix as the observable $\hat{A}$. In such a case, Eq. (\ref{FD_theorem}) becomes
\begin{equation}
\dfrac{\partial}{\partial \gamma} \left\langle \hat{\rho}(\gamma) \right\rangle_{\hat{\rho}(\gamma)}=\left\langle \dfrac{\partial \hat{\rho}(\gamma) }{\partial\gamma} \right\rangle_{\hat{\rho}(\gamma)}+\left\langle \hat{\rho}(\gamma)\dfrac{\partial  \log \hat{\rho}(\gamma) }{\partial\gamma}  \right\rangle_{\hat{\rho}(\gamma)}, \label{QEI_rho_0}
\end{equation}
where the expectation value of the density matrix is the purity of this state, i.e.
\begin{equation}
\left\langle \hat{\rho} \right\rangle_{\hat{\rho}}=\Tr\left\lbrace \hat{\rho}\hat{\rho}\right\rbrace=\varphi.
\end{equation}

Moreover, for this choice of the observable in the QEI, both terms on the right-hand side of Eq. (\ref{QEI_rho_0}) are identical. Indeed:
\begin{equation}
\left\langle \hat{\rho}\dfrac{\partial  \log \hat{\rho} }{\partial\gamma}  \right\rangle_{\hat{\rho}}=\left\langle \hat{\rho}\hat{\rho}^{-1} \dfrac{\partial \hat{\rho}}{\partial \gamma} \right\rangle = \left\langle \dfrac{\partial \hat{\rho} }{\partial\gamma} \right\rangle_{\hat{\rho}}.
\end{equation}
So, depending on the specific form of $\hat{\rho}$, we can recast the QEI for $\hat{A}=\hat{\rho}$ as
\begin{equation}
\dfrac{\partial \varphi (\gamma)}{\partial \gamma} =2\left\langle \dfrac{\partial \hat{\rho}(\gamma) }{\partial\gamma} \right\rangle_{\hat{\rho}(\gamma)}, \label{QEI_purity_1}
\end{equation}
or also as
\begin{equation}
\dfrac{\partial \varphi (\gamma)}{\partial \gamma} =2\left\langle \hat{\rho}(\gamma)\dfrac{\partial  \log \hat{\rho}(\gamma) }{\partial\gamma}  \right\rangle_{\hat{\rho}(\gamma)}. \label{QEI_purity_2}
\end{equation}

In Section \ref{purity_quantum_section} we consider the case $\left(\hat{\rho},\lambda \right) $ for a density matrix representing a mixture of two pure states with $\gamma=\lambda$ (a quantum-like parameter). Then, in Section \ref{purity_beta_section}, we consider the pair $\left(\hat{\rho}_c,\beta \right) $ for a canonical density matrix $\hat{\rho}_c$ and the inverse temperature as the continuous parameter.

\subsection{Dependence of $\varphi$ on a quantum-like parameter} \label{purity_quantum_section}
As a first application of the QEI for $\hat{A}=\rho$, consider a mixture of two single projectors $ \left\lbrace \ket{0}\bra{0},\ket{1}\bra{1}\right\rbrace$ modulated by a bounded parameter $\lambda$, such that $0\leq \lambda \leq 1$. The density matrix for this model is 
\begin{equation}
\hat{\rho}(\lambda)=\left(1-\lambda \right) \ket{0}\bra{0} + \lambda\ket{1}\bra{1}. \label{density_matrix_purity_1}
\end{equation}
Additionally, we assume that the pure states $\left\lbrace \ket{i} \right\rbrace_{i=1,2}$ constitute an orthogonal basis. 
Due to the form of $\hat{\rho}$ in this case, it is convenient to use the QEI in the form given by Eq. (\ref{QEI_purity_1}). Thus, the derivative of $\hat{\rho}(\lambda)$ with respect to $\lambda$ turns out to be
\begin{equation}
\dfrac{\partial \hat{\rho}}{\partial \lambda}=-\ket{0}\bra{0}+\ket{1}\bra{1},
\end{equation} 
so that
\begin{equation}
\begin{split}
\left\langle \dfrac{\partial \hat{\rho} }{\partial\lambda} \right\rangle_{\hat{\rho}} & =  \Tr \left\lbrace \dfrac{\partial \hat{\rho}}{\partial \lambda} \hat{\rho} \right\rbrace \\
& = \Tr \left\lbrace \left[-\ket{0}\bra{0}+\ket{1}\bra{1}   \right] \left[\left(1-\lambda \right) \ket{0}\bra{0} + \lambda\ket{1}\bra{1} \right]  \right\rbrace\\
& = 2\lambda-1. 
\end{split}
\end{equation}
Using the above expectation value, we get to a first-order differential equation for $\varphi$, namely
\begin{equation}
\dfrac{\partial \varphi}{\partial \lambda}=2 (2\lambda-1). \label{purity_edo_lambda}
\end{equation}
When integrating Eq. (\ref{purity_edo_lambda}), we obtain the following solution:
\begin{equation}
\varphi(\lambda)=2\lambda(\lambda-1)+\varphi_0, \label{purity_solution}
\end{equation}
where $\varphi_0$ is an integration constant which is determined when evaluating both $\hat{\rho}(\lambda)$ and $\varphi(\lambda)$ at $\lambda=0$ and $\lambda=1$. As to the density matrix (Eq. (\ref{density_matrix_purity_1})), we have
 \begin{equation}
 \hat{\rho}(\lambda)=
    \begin{cases}
        \ket{0}\bra{0} & \text{if } \lambda=0\\
         \ket{1}\bra{1} & \text{if } \lambda=1
    \end{cases}
\end{equation}
i.e., for both extreme values $\lambda=0$ and $\lambda=1$, we obtain the pure state projectors $\ket{0}\bra{0}$ and $\ket{1}\bra{1}$, respectively. As to the solution for the purity (Eq. \ref{purity_solution}), we have
    \begin{equation}
  \varphi (\lambda)=
    \begin{cases}
        \varphi_0 & \text{if } \lambda=0\\
        \varphi_0 & \text{if } \lambda=1
    \end{cases}
\end{equation}
Because the single projectors $\ket{0}\bra{0}$ and $\ket{1}\bra{1}$ are attached to the extreme values of $\lambda$, the purity $\varphi$ must be equal to $1$ for both values $\lambda=0$ and $\lambda=1$. Therefore, $\varphi_0=1$.

\begin{figure}[htb]
\centering
\includegraphics[width=0.7\linewidth]{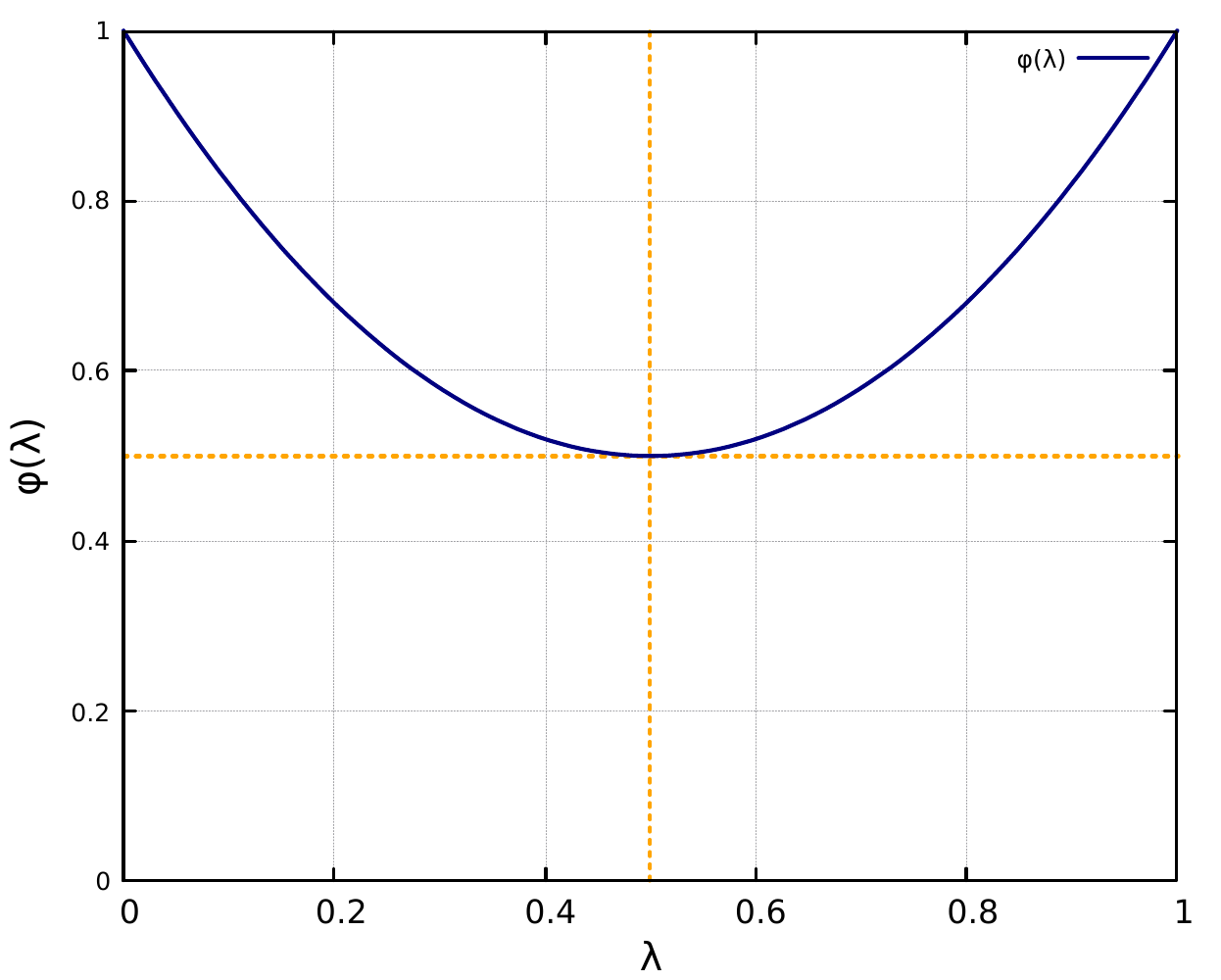}
\caption{Purity for a mixture of two pure states.} \label{fig1}
\end{figure}

Figure \ref{fig1} shows the form of $\varphi$ for this mixture of two single projectors, where it is possible to identify that the minimum of the curve (minimum purity) is reached at $\lambda=1/2$ having a value of $\varphi(\lambda=1/2)=1/2$, which corresponds to the inverse of the dimensionality of the subspace of the Hilbert space in which the density matrix is spanned.

\subsection{Depedence of $\varphi$ on $\beta$} \label{purity_beta_section}

As a second application of the QEI for $\hat{A}=\hat{\rho}$ we consider a canonical density matrix $\hat{\rho}_c$ (Eq. (\ref{canonical_density_matrix})). In this case we take $\gamma=\beta$ and, due to the form of $\hat{\rho}_c$, it is convenient to use the version of the QEI given by Eq. (\ref{QEI_purity_2}). After performing a similar algebraic treatment to that of Section (\ref{canonical_model_section}), we obtain
\begin{equation}
\dfrac{\partial \varphi (\beta)}{\partial \beta}=-2\mathrm{Cov} \left( \hat{\rho},\hat{H} \right). \label{FD_purity_beta}
\end{equation}
We can also recast the above identity in terms of the absolute temperature $T$ instead of $\beta$ obtaining
\begin{equation}
\dfrac{\partial \varphi (T)}{\partial T}=\dfrac{2}{kT^2}\mathrm{Cov} \left( \hat{\rho},\hat{H} \right). \label{FD_purity_T}
\end{equation}
What the above identity tells us is that the dependence of the purity associated with a quantum state $\hat{\rho}$ on the absolute temperature is directly related to the fluctuation of the energy of the system with respect to the thermal state $\hat{\rho}$. Thus, the stronger the covariance between the density matrix with the Hamiltonian, the greater the dependence of the purity on temperature. Hence, we called to the identity Eq. (\ref{FD_purity_T}) the \textit{purity-fluctuation-dissipation theorem (p-FDT)}.

A corollary of Eqs. (\ref{FD_purity_beta}) and (\ref{FD_purity_T}) is the following: because the purity of a pure state is equal to $1$, and thus $\partial \varphi/\partial \beta=0$, we conclude
\begin{equation}
\mathrm{Cov}\left( \hat{\rho},\hat{H} \right)=0. \label{purity_corollary}
\end{equation}
Therefore, the Hamiltonian and the density matrix are uncorrelated variables for a quantum system described by a pure state.

As an illustrative example of the purity-fluctuation-dissipation theorem (in the form of Eq. (\ref{FD_purity_beta})), consider the quantum harmonic oscillator model whose Hamiltonian is given by
\begin{equation}
\hat{H}=\hbar\omega\left(\hat{N}+1/2 \right), 
\end{equation}
where $\omega$ is the angular frequency of the oscillator and $\hat{N}$ is, as usual, the number operator. Then, the canonical density matrix for this quantum model is
\begin{equation}
\hat{\rho}_c(\beta)=\dfrac{\exp{ -\beta\hbar\omega (\hat{N}+1/2 )} }{Z(\beta)}.
\end{equation}
In order to calculate the purity for the above density matrix, we recast $\varphi(\beta)$ as
\begin{equation}
\varphi(\beta)=\Tr\left\lbrace \hat{\rho}^2_c(\beta)\right\rbrace =\Tr\left\lbrace \dfrac{\left( e^{-\beta\hat{H}} \right)^2 }{Z^2(\beta)} \right\rbrace = \dfrac{\Tr \left\lbrace \left( e^{-\beta\hat{H}}\right)^2  \right\rbrace }{\left( \Tr\left\lbrace e^{-\beta\hat{H}} \right\rbrace  \right)^2 }, \label{purity_harmonic_1}
\end{equation}
where
\begin{equation}
Z(\beta)=\Tr\left\lbrace e^{-\beta\hat{H}}\right\rbrace 
\end{equation} 
is the corresponding partition function. On the one hand, when using the spectrum of the harmonic oscillator, which is given by 
\begin{equation}
\varepsilon_n=\hbar\omega(n+1/2), \hspace{0.5cm}  n\in \mathbb{N}_0,
\end{equation}
the numerator of Eq. (\ref{purity_harmonic_1}) becomes 
\begin{equation}
\Tr\left\lbrace e^{-2\beta\hat{H}} \right\rbrace = \left[2\sinh(\beta\hbar\omega) \right]^{-1}. 
\end{equation}
On the other hand, calculating the partition function for this model is a straightforward exercise in a standard statistical mechanics course \cite{greiner}. The result is
\begin{equation}
Z(\beta)=\left[ 2\sinh\left( \dfrac{\beta\hbar\omega}{2} \right)  \right]^{-1}. 
\end{equation}
Therefore, $\varphi(\beta)$, according to Eq. (\ref{purity_harmonic_1}), turns out to be
\begin{equation}
\varphi(\beta)=2\dfrac{\sinh^2\left( \dfrac{\beta\hbar\omega}{2} \right) }{\sinh(\beta\hbar\omega)}.
\end{equation}
In addition, by use of some properties of hyperbolic functions, $\varphi(\beta)$ can also be rewritten as
\begin{equation}
\varphi(\beta)=\tanh\left( \dfrac{\beta\hbar\omega}{2} \right). \label{purity_tanh}
\end{equation}

%\begin{figure}[htb]
%\centering
%\includegraphics[width=0.7\linewidth]{purity_beta.pdf}
%\caption{Purity for a mixture of two pure states}
%\end{figure}

\begin{figure}[htp] 
    \centering
    \subfloat[Behavior of the purity with $\beta$.]{ \label{fig:a}
        \includegraphics[width=0.48\textwidth]{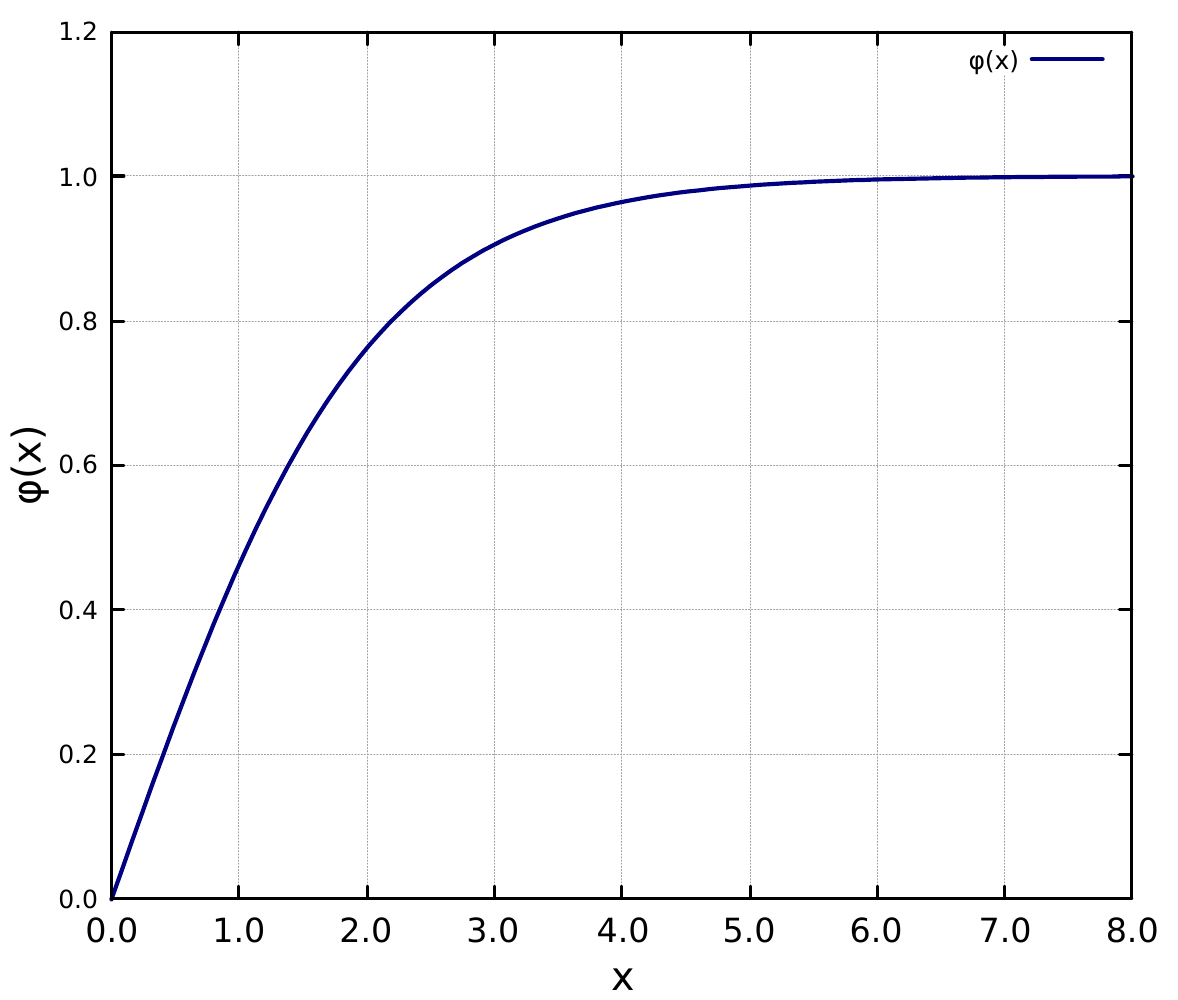}
        
        }
    \hfill
    \subfloat[Covariances between $\hat{H}$ and $\hat{\rho}_c(\beta)$.]{  \label{fig:b}
        \includegraphics[width=0.48\textwidth]{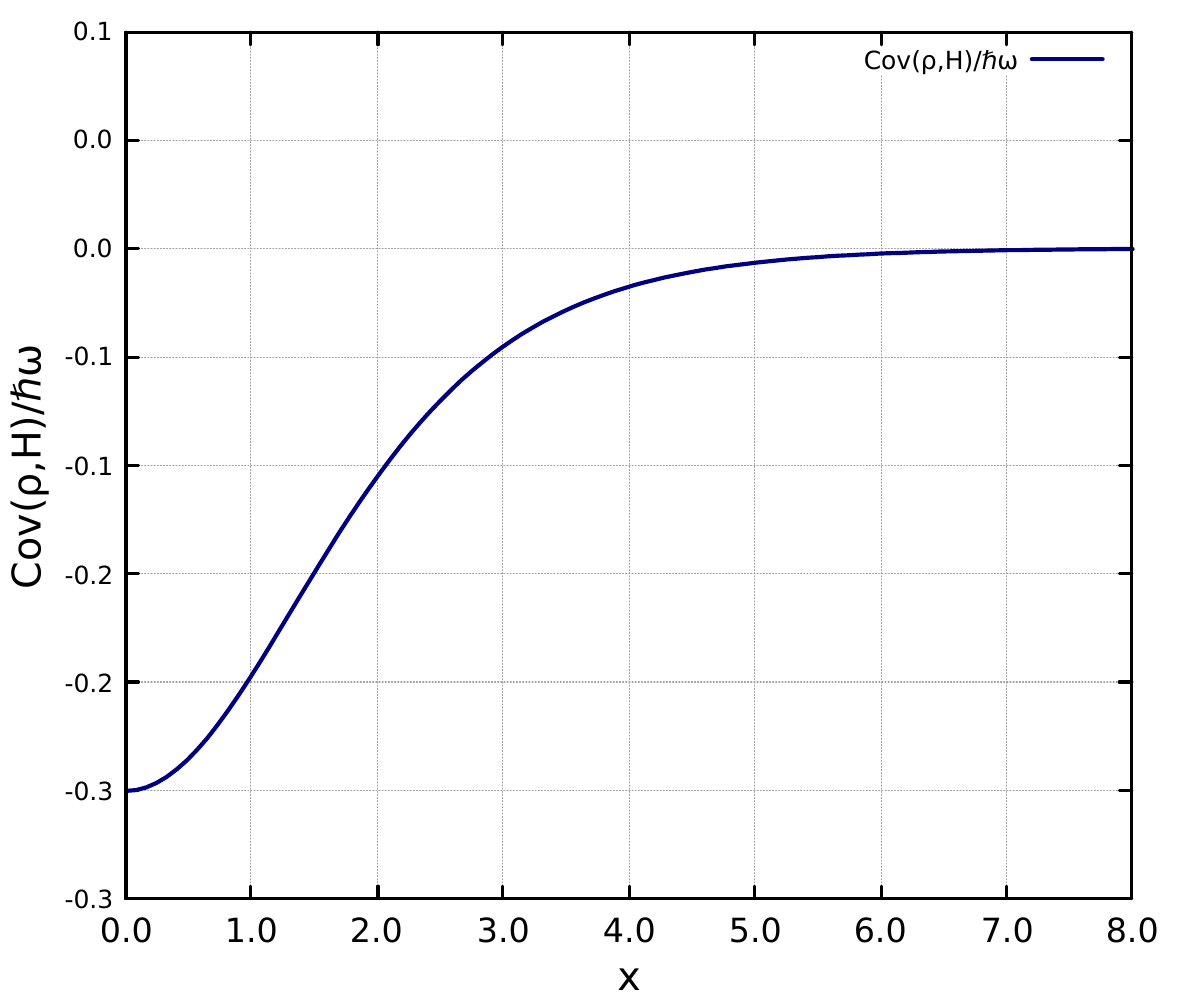}
      
        }
    \caption{Quantum purity and covariances for the harmonic oscillator model in the canonical model. In both graphs, it has been used a dimensionless reduced variable defined by $x\equiv\beta\hbar\omega$. The covariances were obtained by use of the p-FDT (Eq. (\ref{FD_purity_beta})).} \label{fig2}
\end{figure}

The advantage of the p-FDT is that it provides a way of calculating the covariances between the Hamiltonian and the density matrix, by means of the derivative with respect to the inverse temperature of the purity associated with such a density matrix. Thus, from Eq. (\ref{purity_tanh}), $\partial \varphi(\beta)/\partial \beta$  turns out to be
\begin{equation}
\dfrac{\partial \varphi(\beta)}{\partial \beta}=\dfrac{\hbar \omega}{2}\sech^2\left( \dfrac{\beta\hbar\omega}{2}\right), 
\end{equation}
and by comparison with the p-FDT (Eq. (\ref{FD_purity_beta})), we finally obtain an analytic expression for the covariances between $\hat{H}$ and $\hat{\rho}_c(\beta)$:
\begin{equation}
\mathrm{Cov} \left( \hat{\rho},\hat{H} \right)=-\dfrac{\hbar\omega}{4}\sech^2\left( \dfrac{\beta\hbar\omega}{2}\right).
\end{equation}

Figure \ref{fig2} shows both the behavior of the purity associated with a canonical density matrix for the harmonic oscillator model and the corresponding covariances between $\hat{H}$ and $\hat{\rho}_c(\beta)$. Specifically in Fig. \ref{fig:a} we can notice that as the inverse temperature increases (the absolute temperature decreases), the purity tends to its maximum value given by the unity. Thus, in the limit of zero (absolute) temperature, the quantum states are strictly pure.  On the other hand, from Fig. \ref{fig:b} we can notice that as the inverse temperature increases, the covariance between the Hamiltonian and the canonical density matrix tends to zero, which means that in the limit of zero (absolute) temperature $\hat{H}$ and $\hat{\rho}_c(\beta)$ are uncorrelated variables (their fluctuations are independent of each other). The above is in complete accordance with the corollary of the p-FDT stated in Eq. (\ref{purity_corollary}).

\section{Generalized MaxEnt density matrix identities} \label{generalized_section}
Given a set $\left\lbrace \alpha_i \right\rbrace_{i=1,...,n}$ of statistical-like parameters, and another set $\left\lbrace \lambda_i \right\rbrace_{i=1,...,m}$ of quantum-like parameters, we can generalize the results of the previous sections introducing a \textit{generalized} density matrix of the form
\begin{equation}
\hat{\rho}\left( \alpha_1,...,\alpha_n;\lambda_1,...,\lambda_m \right) =\dfrac{1}{\mathcal{Z}\left( \alpha_1,...,\alpha_n;\lambda_1,...,\lambda_m\right) }\exp\left[ -\sum_{j=1}^n \alpha_j \hat{F}_j\left( \lambda_1,...,\lambda_m\right) \right], \label{generalized_DM}
\end{equation}
where $\left\lbrace \hat{F}_j\right\rbrace_{j=1,...,n}$ is a set of observables acting on the Hilbert space $\mathcal{H}$ of states of the system (e.g. Hamiltonian, number operator, etc.). These observables are compatible with each other and may depend on one or more quantum-like parameters. It is important to notice that the statistical-like parameters depend on the statistical model only and appear as Lagrange multipliers associated with certain constraints in the MaxEnt procedure, while the quantum-like parameters are related to the particular features of the system, and some of them can be contained in the Hamiltonian or other observables of the system (e.g., perturbation parameters). We have used the appellative \textit{generalized} for the density matrix in the sense given in Ref. \cite{jaynes1}: within the approach of MaxEnt inference not only does it have the Hamiltonian the prime role, but all the observables in the set $\left\lbrace \hat{F}_j\right\rbrace_{j=1,...,n}$ possess equal weight.

The entropy associated with such a generalized density matrix is obtained through the von-Neumann entropy (\ref{von_neumann_entropy}), which in this case is given by
\begin{equation}
S= k\sum_{j=1}^n \alpha_j \left\langle \hat{F}_j \right\rangle_{\hat{\rho}}+k\ln\mathcal{Z}\left( \alpha_1,...,\alpha_n;\lambda_1,...,\lambda_m\right). \label{entropy}
\end{equation}
We can easily identify the canonical and the grand-canonical cases encoded in the entropy Eq. (\ref{entropy}). For instance, considering only one statistical-like parameter $\alpha$ and setting it equal to the inverse temperature $\beta$, and choosing $\hat{F}=\hat{H}$, we obtain 
\begin{equation}
-kT\ln \mathcal{Z}(\beta;\lambda_1,...,\lambda_m)=E-TS=\mathcal{F}
\end{equation}
where $\mathcal{F}$ is the Helmholtz free energy (\ref{free_energy}). On the other hand, if we consider two statistical-like parameters,  namely, $\alpha_1=\beta$ and $\alpha_2=-\beta\mu$, and setting $\hat{F}_1=\hat{H}$ and $\hat{F}_2=\hat{N}$ for the corresponding observables, we can recover the grand-canonical equation of state
\begin{equation}
\ln \Xi(\beta,\mu;\lambda_1,...,\lambda_m)=\dfrac{pV}{kT}=-\dfrac{\Phi}{kT},
\end{equation}
where $\Xi(\beta,\mu;\lambda_1,...,\lambda_m)$ and $\Phi$ are the grand partition function and the grand potential, respectively.

In order to obtain generalized quantum expectation identities using the MaxEnt density matrix, it is necessary to bear in mind the different choices for the continuous parameter $\gamma_k$ in the QEI, i.e., $\gamma_k \in \left\lbrace \alpha_1,...,\alpha_n \right\rbrace \cup \left\lbrace  \lambda_1,...,\lambda_m  \right\rbrace $, with $k \in \left\lbrace 1,2,...,n,n+1,...,n+m\right\rbrace$. In this way we have the following ordering of parameters:
\begin{equation}
\gamma_1=\alpha_1,\gamma_2=\alpha_2,...,\gamma_n=\alpha_n,\gamma_{n+1}=\lambda_1,...,\gamma_{n+m}=\lambda_m.
\end{equation}

Using the model given by (\ref{generalized_DM}) in ($\ref{FD_theorem}$) we obtain the QEI in its generalized form:

\begin{equation}
\dfrac{\partial}{\partial \gamma_k} \left\langle \hat{A}\right\rangle_{\hat{\rho}}=\left\langle \dfrac{\partial \hat{A} }{\partial\gamma_k} \right\rangle_{\hat{\rho}}-\sum_{j=1}^n\left\langle \hat{A} \dfrac{\partial}{\partial\gamma_k} \left( \alpha_j\hat{F}_j  \right)    \right\rangle_{\hat{\rho}}-   \dfrac{\partial \ln \mathcal{Z}}{\partial \gamma_k}   \left\langle  \hat{A}  \right\rangle_{\hat{\rho}}, \label{generalized-Q-FDT}
\end{equation}
where $\hat{A}$ is an observable depending on quantum-like parameters only and must be compatible with the generalized density matrix $\hat{\rho}$.
As in the previous section, we can choose different operators for the observable $\hat{A}$. The results for $\hat{A}=\mathds{1}$ and $\hat{A}=\hat{F}_k$, for both types of continuous parameters, are summarized in Table \ref{table_3}. These four cases encode all the possibilities for the QEI. As a result of introducing a generalized density matrix, we have gained breath and depth: breadth because we can see all the above expectation identities, using both the canonical and grand canonical models, as particular instances of Eq. (\ref{generalized-Q-FDT}); depth by being able to frame several quantum-statistical identities within an inferential-statistic (and Bayesian) framework. 

Consider the first row in Table \ref{table_3} for $\hat{A}=\mathds{1}$. We have for the pairs $\left(\mathds{1},\lambda_k \right)$ and $\left(\mathds{1},\alpha_k \right)$ the identities
\begin{equation}
\sum_{j=1}^n \alpha_j \left\langle \dfrac{\partial \hat{F}_j}{\partial \lambda_k} \right\rangle_{\hat{\rho}}=-\dfrac{\partial \ln\mathcal{Z}}{\partial \lambda_k} \label{gemeralized_1}
\end{equation}
and
\begin{equation}
\left\langle \hat{F}_k  \right\rangle_{\hat{\rho}}=-\dfrac{\partial \ln \mathcal{Z}}{\partial \alpha_k}, \label{generalized_2}
\end{equation}
where Eq. (\ref{gemeralized_1}) is a generalization of the thermodynamic integration formula for $n$ Lagrange multipliers, while Eq. (\ref{generalized_2}) establishes a relation between the mean value of the $kth$ observable of the set with its respective conjugate variable (parameter). It is worth mentioning that Eq. (\ref{generalized_2}) agrees with the identity previously obtained, by means of maximum-entropy inference by E. T. Jaynes in the reference \cite{jaynes2}. That is how our approach differs from previous approaches: not only are we using a density matrix coming from a MaxEnt procedure, but we are also systematizing all the expectation identities coming from the canonical and the grand canonical models within the same general theorem (Theorem (\ref{theorem_1})).

Finally, for the pairs $(\hat{F}_k, \lambda_l)$ and $(\hat{F}_k,\alpha_l)$ in the second row of Table \ref{table_3} we have the following identities:
\begin{equation}
\dfrac{\partial }{\partial \lambda_l}\left\langle \hat{F}_k \right\rangle_{\hat{\rho}} = \left\langle   \dfrac{\partial \hat{F}_k}{\partial \lambda_l} \right\rangle_{\hat{\rho}}-\sum_{j=1}^n \alpha_j\hspace{0.06cm}\mathrm{Cov} \left(  \hat{F}_k, \dfrac{\partial \hat{F}_j}{\partial \lambda_l}  \right)\label{generalized_3}
\end{equation}
and
\begin{equation}
\dfrac{\partial}{\partial \alpha_l}\left\langle \hat{F}_j \right\rangle_{\hat{\rho}}=-\mathrm{Cov}\left( \hat{F}_j,\hat{F}_l\right),\label{generalized_4}
\end{equation}
where 
\begin{equation}
\begin{split}
\mathrm{Cov} \left(  \hat{F}_k,\hat{F}_l\right)  & \equiv \left\langle \Delta \hat{F}_k\Delta \hat{F}_l\right\rangle_{\hat{\rho}} \\
&=\left\langle  \hat{F}_k \hat{F}_l \right\rangle_{\hat{\rho}}-\left\langle \hat{F}_k \right\rangle_{\hat{\rho}}\left\langle \hat{F}_l \right\rangle_{\hat{\rho}}
\end{split}
\end{equation}
are the covariances between the observables $\hat{F}_k$ and $\hat{F}_l$ in the quantum state $\hat{\rho}$. We observe that Eq. (\ref{generalized_3}) is the generalization of the identity (\ref{canonical_H_lambda}) for $n$ different Lagrange multipliers while, Eq. (\ref{generalized_4}) sets up a generalization of the fluctuation-dissipation-like identities (cf. Eqs (\ref{canonical4}) for the canonical model, and Eqs. (\ref{grand_canonical_H_mu}), (\ref{grand_canonical_N_beta}) and (\ref{grand_canonical_N_mu}) for the grand canonical model).

%\begin{center}
\begin{table}
\begin{footnotesize}
\begin{tabular}{| c || c | c  |}
\hline 
\textbf{Operator} & \color{blue}$\gamma_k=\lambda_k$ \color{black}(quantum-like parameter) & \color{red}$\gamma_k=\alpha_k$ \color{black}(statistical-like parameter) \\ \hline                        
$\hat{A}=\mathds{1}$ & $\sum_{j=1}^n \alpha_j \left\langle \dfrac{\partial \hat{F}_j}{\partial \lambda_k} \right\rangle_{\hat{\rho}}=-\dfrac{\partial \ln\mathcal{Z}}{\partial \lambda_k}$ &  $\left\langle \hat{F}_k  \right\rangle_{\hat{\rho}}=-\dfrac{\partial \ln \mathcal{Z}}{\partial \alpha_k}$ \\
$\hat{A}=\hat{F}_k$ & $\dfrac{\partial }{\partial \lambda_l}\left\langle \hat{F}_k \right\rangle_{\hat{\rho}} = \left\langle   \dfrac{\partial \hat{F}_k}{\partial \lambda_l} \right\rangle_{\hat{\rho}}-\sum_{j=1}^n \alpha_j\hspace{0.06cm}\mathrm{Cov} \left(  \hat{F}_k, \dfrac{\partial \hat{F}_j}{\partial \lambda_l}  \right)$  & $\dfrac{\partial}{\partial \alpha_l}\left\langle \hat{F}_j \right\rangle_{\hat{\rho}}=-\mathrm{Cov}\left( \hat{F}_j,\hat{F}_l\right)$  \\
\hline  
\end{tabular}
\caption{Generalized quantum expectation identities. Depending on the type of the continuous parameter $\gamma_k$ and the choice of the observable, we obtain four different cases.} \label{table_3}
\end{footnotesize}
\end{table}
%\end{center}

%----------------------------------------------------------------------------------------------------------------------------

%\begin{equation}
%\begin{split}
%\Tr \left\lbrace \hat{A} \left[ \hat{H},\hat{\rho} \right] \right\rbrace & = \Tr \left(  \hat{A} \hat{H} \hat{\rho} \right) -\Tr \left(  \hat {A} \hat{\rho} \hat{H} \right) ,\\
%&  = \Tr \left(  \hat{A} \hat{H} \hat{\rho} \right)  -\Tr \left(  \hat {H} \hat{A} \hat{\rho} \right) ,\\
%& = \Tr \left\lbrace \left[ \hat{A},\hat{H} \right] \hat{\rho}  \right\rbrace,\\
%& = \left\langle \left[ \hat{A},\hat{H}\right]  \right\rangle_{\hat{\rho}(t)}, 
%\end{split}
%\end{equation}

%= \left\langle \hat{F}_k\dfrac{\partial}{\partial \alpha_l} \left[ -\sum_{j=1}^n  \alpha_j \hat{F}_j -\ln \mathcal{Z} \hspace{0.05cm}%\mathds{1}  \right]  \right\rangle_{\hat{\rho}}, \\
%&

\section{Concluding remarks}
In this work we derived a quantum expectation identity (Eq. (\ref{FD_theorem})) coming from Bayesian statistics in the language and rules of quantum mechanics, and we exposed the usefulness of this, by including several theorems of quantum mechanics and quantum statistics, among others, \textit{beneath its shade}.

A prospective and interesting challenge would be exploring the scope of another kind of expectation identities, such as the \textit{conjugate-variables} identities, in quantum statistics. In this sense, we would have a complete program of Bayesian expectation identities in the context of quantum statistical mechanics.

\section{Acknowledgments}

\hspace{0.7cm} Ms. Vania S\'aez is thanked for her advice and encouragement, and also for her inspiring ideas concerning the nature of matter and \textit{time}. Ms. Yasna Ortiz is thanked for her grammatical advice and her assistance in the preparation of the speech given in the IWOSP 2023 conference. Prof. R. C. Bochicchio is thanked for his time dedicated to interesting conversations about the relationship between molecular physics and statistical physics. 

SD and BM thankfully acknowledge funding from ANID FONDECYT 1220651 grant. 

By last, DP and BM’s research has been partially supported by ANID Chile via the project FONDECYT N° 1241719.

%    ^..^      /
%    /_/\_____/
%      /\   /\
%  	  /  \ /  \

%
%
% Miss Guadalupe and miss Andina are thanked for their silent and constant companion. 
\section*{References}

\bibliography{q_fdt}
\bibliographystyle{elsarticle-num}

 \renewcommand{\theequation}{A-\arabic{equation}}
  % redefine the command that creates the equation no.
  \setcounter{equation}{0}  % reset counter 

\section*{Appendix: Derivative of the logarithm of an operator} \label{appendix}
Consider a linear operator $\hat{\rho}(\gamma)$ dependent on a continuous parameter $\gamma$. Then, the Taylor series of the natural logarithm of $\hat{\rho}(\gamma)$ is given by \cite{lie_groups, BHC_so3}
\begin{equation}
\log \hat{\rho}(\gamma) = -\sum_{k=1}^{\infty} \dfrac{\left( \mathds{1}-\hat{\rho}(\gamma)\right)^k }{k}.
\end{equation}
It is worth mentioning that the above power series is only valid for \textit{bounded operators} \cite{functional}. Fortunately, this is the case of the density matrix $\hat{\rho}(\gamma)$. Taking the derivative of $\log \hat{\rho}(\gamma)$ with respect to $\gamma$ we have

\begin{equation}
\begin{split}
\dfrac{\partial \log \hat{\rho}(\gamma)}{\partial \gamma} & =-\sum_{k=1}^{\infty} \left( \mathds{1}-\hat{\rho}(\gamma) \right)^{k-1}\cdot -\dfrac{\partial \hat{\rho}(\gamma)}{\partial \gamma}\\ 
& = \sum_{j=0}^{\infty} \left( \mathds{1}-\hat{\rho}(\gamma) \right)^{j}\dfrac{\partial \hat{\rho}(\gamma)}{\partial \gamma}, \label{appendix_1}
\end{split}
\end{equation}
where we have relabeled the sum taking $j=k-1$. On the other hand, from the geometric series we can write the expansion of the inverse of $\hat{\rho}(\gamma)$, namely
\begin{equation}
\left[ \mathds{1}- \left( \mathds{1}-\hat{\rho}(\gamma) \right)  \right]^{-1}=\sum_{j=0}^{\infty} \left(\mathds{1}-\hat{\rho}(\gamma) \right)^{j}=\hat{\rho}^{-1}(\gamma),  \label{appendix_2}
\end{equation} 
where the above power series is also valid for bounded operators only. Hence, when using Eq. (\ref{appendix_2}) in Eq. (\ref{appendix_1}) we finally obtain the desired equality
\begin{equation}
\dfrac{\partial \log \hat{\rho}(\gamma)}{\partial \gamma}=\hat{\rho}^{-1}(\gamma)\dfrac{\partial \hat{\rho}(\gamma)}{\partial \gamma}.
\end{equation}

\end{document}